\documentclass[11pt]{article}
\usepackage[T1]{fontenc}
\usepackage[utf8]{inputenc}
\usepackage{mathpazo}
\usepackage[margin=1in]{geometry}
\usepackage{amsmath,amssymb,amsthm}
\usepackage{graphicx}
\usepackage{float}
\usepackage{booktabs}
\usepackage{multirow}
\usepackage{array}
\usepackage{xcolor}
\usepackage[labelfont=bf]{caption}
\usepackage[shortlabels]{enumitem}
\usepackage{microtype}
\usepackage{physics}
\usepackage{quantikz}
\usepackage{bm}
\usepackage{algorithm}
\usepackage{algorithmic}
\usepackage{tikz}
\usetikzlibrary{arrows.meta, positioning, shapes.geometric, calc}
\usepackage[numbers,sort&compress]{natbib}
\usepackage{authblk}
\usepackage[colorlinks=true,linkcolor=blue!60!black,citecolor=blue!60!black,urlcolor=blue!60!black]{hyperref}

\title{Hybrid Quantum Algorithms for Trade Surveillance Analytics}
\author[1,$\dagger$,*]{Bimal Mehta}
\author[1,$\dagger$,*]{Daniel musgrove}
\author[2,$\dagger$]{Devanshu Shekhar}
\author[2,$\dagger$]{Shiplu Sarkar}
\author[2,$\dagger$]{Vanshaj Kerni}
\author[2,$\dagger$,*]{Varun Puram}
\author[2]{Nitesh Kansara}
\author[2]{Abinash Pati}
\author[3]{Sarthak Chakraborty}
\author[3]{Parth Gaglani}
\author[2]{Bhaskar Roy Bardhan}
\affil[1]{Vanguard}
\affil[2]{The Washington Institute for STEM, Entrepreneurship, and Research (WISER), Washington DC, USA}
\affil[3]{Capgemini}
\affil[$\dagger$]{These authors contributed equally to this work.}
\affil[*]{Correspondence: \href{mailto:bimal_mehta@vanguard.com}{bimal\_mehta@vanguard.com} , \href{mailto:daniel_musgrove@vanguard.com}{daniel\_musgrove@vanguard.com}, \href{mailto:varun.p@thewiser.org}{varun.p@thewiser.org}}
\date{}

\begin{document}
\maketitle

\begin{abstract}
Anti-money laundering (AML) and trade-surveillance work today is still largely manual: rule-based monitoring systems raise alerts of which, by industry estimates, more than $95\%$ are false positives, and every one of them is reviewed by a human analyst. This paper benchmarks four families of quantum and hybrid quantum--classical classifiers against classical baselines on a synthetic pump-and-dump detection task: a fidelity-based quantum $k$-nearest-neighbours classifier (QKNN) with clustering-based neighbourhood routing; a quantum support vector machine (QSVM) built on a one-body projected quantum kernel; a quantum neural network (QNN) with a second-order Pauli-ZZ feature map and a RealAmplitudes ansatz on $20$ qubits; and a hybrid quantum neural network (HQNN) in which a small classical feed-forward network generates every rotation angle of a five-qubit circuit. The classical baselines are logistic regression, XGBoost, one- and three-hidden-layer neural networks, and a classical $k$-nearest-neighbours control. All models share one stratified split of $1000$ accounts ($800$ training, $200$ test, $20\%$ suspicious) and two $20$-feature representations (ANOVA feature selection and PCA); the QKNN and QSVM are also evaluated under simulated device noise, and the trained HQNN is executed on the \texttt{ibm\_fez} superconducting processor. The strongest model is classical, XGBoost on ANOVA features, at a suspicious-class F1 of $61.4\%$. The HQNN is the strongest quantum-hybrid model, reaching $54.5$--$57.9\%$ F1 on an ideal simulator and $55.4$--$59.8\%$ on \texttt{ibm\_fez}, ahead of six of the eight classical configurations; at a $15\%$ false-positive rate, one in every two of its alerts is a genuinely suspicious account. Compared with the $20$-qubit QNN ($42.0\%$ F1), it raises fewer than half as many false alarms at a similar detection rate while using a quarter of the qubits. The QKNN reaches $43.5\%$ F1 ($44.4\%$ under noise), above its classical counterpart ($35.6\%$), and the QSVM does not exceed $41\%$. We do not observe a quantum advantage on this benchmark, but the results point to considerable potential: a hybrid model on only five qubits, run on today's noisy hardware, already performs on par with well-tuned classical models, and richer circuits, larger devices and better training leave substantial room for quantum approaches to improve trade-surveillance analytics. The contribution of this work is a like-for-like comparison of four quantum classifier families against classical controls on shared data, splits and representations.

\end{abstract}

\section{Introduction}
\label{sec:introduction}
\subsection{Motivation}
The global financial system loses an estimated two to five percent of global GDP to money laundering each year, representing a staggering \$800 billion to \$2 trillion annually \cite{unodc2011}.
Despite decades of regulatory tightening and massive investments in compliance technology, the vast majority of laundered funds continue to evade detection \cite{pol2020}.
Anti money laundering (AML) efforts today are still reliant on rule based transaction monitoring systems that generate alerts when transactions breach predefined thresholds, match sanctioned entity lists, or follow rigid behavioural templates.
In practice, such systems produce false positive rates that often exceed 95\%, burying compliance teams in alerts and forcing manual investigation of a large volume of benign activity \cite{han2021, lexisnexis2022}.
Trade surveillance the monitoring of trading venues for pump and dump, insider dealing, spoofing, and other abusive practices is a critical and complementary pillar of the AML framework, sharing the same data intensive, cross product, and cross market monitoring challenges \cite{fatf2021}.
Like AML transaction monitoring, trade surveillance systems predominantly operate on static rules and simple statistical thresholds, which struggle to capture the evolving patterns exploited by sophisticated actors who deliberately structure their trading to mimic legitimate strategies \cite{fincen2021}.
Regulatory mandates have further expanded the scope of both AML and trade surveillance, most notably with the European Union's Sixth Anti Money Laundering Directive (AMLD6) and the priorities set forth by the U.S. Financial Crimes Enforcement Network (FinCEN), which increasingly hold institutions accountable for failing to detect complex schemes that span asset classes and jurisdictions \cite{amld6}.
This tightening landscape, combined with the exponential growth of transactional data has pushed classical monitoring approaches to their practical limits, requiring new computational paradigms that can operate over high dimensional and relational patterns without high false alarm rates.
\subsection{Why Classical Machine Learning Struggles}
Even when machine learning replaces static rules, the nature of financial crime data imposes severe obstacles.
The most immediate is extreme class imbalance: illicit transactions represent a tiny fraction often well below $0.1\%$ of all activity, causing standard classifiers to be overwhelmed by legitimate examples and to treat fraud as noise \cite{dalpozzolo2015}.
A second class of challenges arises from the high dimensionality and relational nature of the data.
Transaction records, enriched with counterparty details, geolocation, timestamp patterns, and product-type information, quickly yield thousands of features.
Kernel-based methods, which can capture subtle non-linear interactions, suffer from quadratic scaling in the number of samples rendering exact kernel machines impractical for the millions of transactions processed daily in modern AML systems \cite{rahimi2007}.
Moreover, money laundering and market manipulation are inherently relational: the behavior of one entity only makes sense in the context of its network of counterparties.
Classical machine learning models that treat observations as independent vectors struggle to exploit this graph structure without extensive feature engineering, while graph-aware methods often bring their own scalability bottlenecks \cite{akoglu2015}.
Fraud patterns do not stand still; they evolve in deliberate, adversarial response to detection measures.
Launderers continually adapt, including splitting transfers, adjusting timings, layering through shell companies so that the very distribution of features shifts over time, a phenomenon known as concept drift \cite{dalpozzolo2015}.
Trade surveillance faces similar adversarial dynamics, as manipulative trading algorithms mimic legitimate strategies to stay below alert thresholds.
Static models trained once on historical data rapidly become obsolete, yet updating them continuously on streaming, imbalanced, and partially labelled data remains a costly problem.
A further difficulty is the manual effort that classical pipelines demand before a model ever sees the data.
Raw AML and trade-surveillance inputs are a mix of categorical codes, free-text fields, numerical amounts, timestamps, and relational links between accounts and counterparties.
Domain experts spend enormous effort engineering features that capture velocity spikes, round-amount patterns, unusual counterparty clusters, or cyclical activity \cite{whitrow2009}.
This process is not only expensive and sensitive to adversarial drift, but it also freezes in a limited set of human-recognisable patterns and misses subtle interactions that are not encoded by the analysts.
Efforts to learn representations directly from raw transactional graphs, via autoencoders or graph neural networks, could be useful, yet they still struggle with the continuous transaction processing (often at high velocity), heavily imbalanced, and partially labelled regime that characterises financial crime detection.
The feature-engineering bottleneck thus remains a major impediment, tying the performance of classical systems to the effort made by the human teams who engineer their inputs. These obstacles motivate the exploration of quantum feature maps and quantum kernel methods pursued in this work, which are developed in Sections~\ref{sec:related work}--\ref{sec:methods}.

\subsection{Scope of This Study}
This study compares four quantum or hybrid quantum--classical classifier families against classical baselines on one shared benchmark. The quantum side comprises (i) a fidelity-based quantum $k$-nearest-neighbours classifier (QKNN) with clustering-based neighbourhood routing (Sec.~\ref{sec:methods:qknn}); (ii) a quantum support vector machine (QSVM) built on a one-body projected quantum kernel (Sec.~\ref{sec:methods:qsvm}); (iii) a quantum neural network (QNN) with a fixed second-order Pauli-ZZ feature map and a trained RealAmplitudes ansatz (Sec.~\ref{sec:methods:hqnn}); and (iv) a hybrid quantum neural network (HQNN) in which a classical feed-forward network generates every rotation angle of a five-qubit circuit (Sec.~\ref{sec:hqnn}). The classical side comprises logistic regression, XGBoost, and one- and three-hidden-layer neural networks (Sec.~\ref{sec:methods:classical}), plus a classical $k$-nearest-neighbours control for the QKNN.

Every model is trained on the same $800$ accounts and scored on the same $200$ held-out accounts, using one of two $20$-feature representations of the same raw data (ANOVA feature selection and PCA; Sec.~\ref{sec:dataset}). The QKNN and QSVM are evaluated both on an ideal simulator and under the noise models of IBM fake backends; the QNN and HQNN are trained on an ideal simulator, and the trained HQNN is additionally executed on the \texttt{ibm\_fez} superconducting processor.
% =====================================================================
\section{Related Work}
\label{sec:related work}
\subsection{Classical Approaches to AML and Trade Surveillance}
The earliest and still most widely deployed line of defence consists of rule-based systems.
These range from simple threshold checks flagging transactions above \$10,000, matching entries against sanctions and politically exposed person (PEP) lists to more statistical rules such as Benford's Law tests for unnatural digit distributions and peer-group deviation analysis \cite{han2021,fatf2021}.
Rule engines are trivially interpretable and easy to audit, which explains their regulatory preference.
Their limitation, however, is apparent: static rules cannot adapt to changing criminal behaviour and generate an overwhelming volume of false positives, forcing human investigators to manually review vast numbers of legitimate transactions.
When institutions move beyond rules, the first step is often a shallow supervised learner.
Logistic regression and linear discriminant analysis have been used for scoring the risk of individual transactions or customers, offering transparency and low computational cost \cite{ngai2011}.
More recently, gradient-boosted tree ensembles exemplified by XGBoost, LightGBM, and CatBoost have become the de facto standard in industry benchmarks and Kaggle-style AML competitions, as they automatically capture non-linearities and feature interactions while still providing reasonable training speed on tabular data \cite{chen2016}.
Nonetheless, both logistic models and tree ensembles remain heavily dependent on the same feature engineering pipeline that rule-based systems rely on; they treat transactions as independent rows in a table and may struggle to learn from the relational topology of financial networks without explicit graph-derived features.
Classic unsupervised anomaly detection offers an appealing alternative, particularly given the scarcity of reliable labels.
Isolation Forest and one-class support vector machines (OC-SVM) learn a compact description of ``normal'' behaviour and flag transactions that fall outside this envelope, without needing labelled fraud examples \cite{liu2008,scholkopf2001}.
However, they are sensitive to the feature representation and tend to produce high false-positive rates when normal behaviour is itself diverse and high-dimensional; moreover, they cannot distinguish between genuinely criminal outliers and unusual-but-legitimate activity unless combined with expert feedback.
Deep learning models promise to reduce the feature-engineering burden by learning representations directly from raw or lightly pre-processed data.
Autoencoders, in particular, have been explored for unsupervised fraud detection by reconstructing normal transactions and flagging those with high reconstruction error as anomalies \cite{zhou2017}.
At the same time, the sequential nature of both payment streams and order-book data has motivated temporal architectures: long short-term memory (LSTM) networks and, more recently, Transformer-based models have been applied to model the order and timing of events, detecting suspicious sequences such as layering or spoofing \cite{chalapathy2019}.
While deep models can capture richer patterns than shallow alternatives, they demand large, clean training sets a condition violated by the extreme imbalance and label noise of financial crime data and their opacity complicates regulatory justification.
Graph-based methods directly address the relational essence of money laundering and market manipulation.
Techniques range from classical graph-anomaly detection (egonet features, community-outlier algorithms) to modern graph neural networks (GNNs) that learn node embeddings encoding both local and neighbourhood structure \cite{akoglu2015,weber2019}.
By representing accounts or traders as nodes and transactions or messages as edges, these models can uncover suspicious structures such as tightly connected rings, rapid fund-layering chains, or synchronised trading clusters.
Their principal drawback is scalability: exact graph algorithms are computationally expensive on networks with millions of nodes and edges, and GNN training requires careful handling of the same label-scarce, imbalanced setting that troubles all supervised approaches.
Moreover, interpreting why a particular subgraph was flagged is a challenge, making it difficult to translate graph-model outputs into actionable alerts.
This work benchmarks four classical baselines, logistic regression, gradient-boosted trees (XGBoost) and two feed-forward neural networks, against four quantum classifier families, quantum $k$-nearest neighbours, a quantum support vector machine, a quantum neural network and a hybrid quantum neural network, all described in Section~\ref{sec:methods}.
% =====================================================================
\section{Dataset and Generation}
\label{sec:dataset}
\subsection{Why Synthetic Data}
Real world financial transaction data, especially for AML and trade surveillance, is notoriously difficult to obtain for research and model development.
Strict privacy regulations most prominently the GDPR in Europe and bank secrecy obligations prevent financial institutions from sharing customer records, even in anonymised form \cite{fatf2021}.
As a result, publicly available AML datasets are virtually non-existent.
The few industry collaborations that release data must strip away counterparty identifiers and temporal precision, removing exactly the relational and sequential structure that modern detection models need.
Synthetic data generators sidestep these barriers by producing artificial transactions that mimic the statistical properties of real payment or trading networks without exposing any genuine customer information.
Beyond circumventing privacy constraints, synthetic data provides researchers with something that real-world data cannot: full control over the underlying fraud scenarios.
Ground-truth labels are known with certainty, eliminating the noise and partiality that impact supervised learning from SAR-derived targets.
This allows rigorous benchmarking, reproducible evaluation, and fair comparison across methods conditions that are practically impossible with proprietary or incomplete real datasets \cite{lopezrojas2016}.
Moreover, synthetic generators can deliberately over-sample rare and emerging money-laundering typologies such as trade-based laundering, thus counteracting the extreme class imbalance that otherwise drowns minority classes.
They can also inject patterns that have not yet appeared in historical archives but are plausible under future regulatory or adversarial shifts, enabling stress-testing of detection systems against novel threats.
For all these reasons, synthetic data has become an indispensable tool for exploring new detection paradigms.
\subsection{Pump and Dump Scenario Design}
A pump-and-dump scheme is a form of market manipulation in which a group of actors artificially inflates the price of a security through coordinated, often deceptive, buying activity (the ``pump''), then sells its holdings at the elevated price to unsuspecting investors (the ``dump''), leaving the victims with rapidly depreciating assets.
The synthetic data generator developed for this work operationalises this lifecycle to produce non-trivial patterns that challenge simple rule-based detectors.
Each event follows three contiguous phases: a quiet accumulation period in which the orchestrator buys gradually, driving a slow linear price rise; a pump phase where the orchestrator and recruited helpers buy aggressively, causing an exponential price run-up; and a dump phase in which the orchestrator unloads its position while victims enter, producing an exponential price decay.
In our approach, events can span between one and fourteen days, with one special one-day ``flash pump''.
This variable duration ranging from a few hours to two weeks forces detection models to recognise manipulation signals on multiple timescales.
To mirror real-world collusion, the generator assigns distinct behavioural roles to the suspicious accounts involved in a scheme.
A small set of orchestrators directs the operation; a larger cohort of helpers amplifies the pump; and the remaining suspicious accounts serve as victims who buy near the peak and later sell at a loss.
Each role follows prescribed trade-count profiles and buy-sell mixes, but even suspicious actors conduct a substantial portion of their activity in unrelated securities noise trades to camouflage their involvement.
On top of this, several obscuration layers are injected deliberately to avoid trivial detection.
Decoy accounts blur the line between genuine and manipulative behaviour: a fraction of truly suspicious accounts trade only a minimal number of low-priced stocks and exhibit break-even profit-loss profiles, while a subset of benign accounts are made to mimic suspicious patterns through elevated low-priced-stock proportions and temporal clustering of those trades.
The gap in the proportion of low-priced trades between benign and suspicious accounts is kept intentionally narrow.
The overall framework is fully configurable, with controllable parameters governing the number of benign and suspicious accounts, the number of independent pump-and-dump events, and a transaction density multiplier (for managing per-account transaction numbers).
\subsection{Feature Set and Label Generation}
For each account, a rich collection of behavioural features is derived from the transaction records across multiple rolling time windows.
Windows span intraday horizons (2, 4, 6, and 12 hours) and daily horizons (1, 2, and 5 days), so that each base feature is repeated across time scales, producing a high-dimensional representation that captures manipulation patterns operating at different velocities.
The features are organised into thematic groups that cover various aspects of market abuse.
Trade-size anomaly scores use robust z-scores (historical median and median absolute deviation from a 21-day lookback) to flag buys and sells that are unusually large for a given account.
Wash-trading indicators count matched buy-sell round trips per security and detect rapid flips completed within one hour, while a price-impact metric measures the magnitude of price changes between consecutive trades.
A profit-and-loss ratio and a pattern counter that detects ``buy followed by multiple sells'' sequences directly target the hallmark dynamics of pump-and-dump behaviour.
Temporal regularity is assessed through inter-transaction interval statistics coefficient of variation, Hopkins statistic, mean absolute deviation and intraday entropy, allowing the model to distinguish bursty, manually orchestrated trading from steady automated execution.
Risk-exposure features capture the concentration and proportion of trading in low-priced stocks, and entropy measures over security types and type-to-type transitions reveal rigid, repetitive strategies.
Finally, a co-trading graph built from shared activity in high-risk (low-priced) securities yields network features co-trading degree, community size, edge density that capture coordinated behaviour across accounts.
The labelling protocol follows the synthetic generation design: any account that participates in a pump-and-dump event as orchestrator, helper, or victim receives a Suspicious label, while all remaining accounts are labelled Benign.
The resulting class distribution is usually highly skewed: suspicious accounts constitute only a small minority of the total, mirroring the extreme rarity of manipulative activity in production systems.
Such imbalance makes conventional accuracy misleading and requires evaluation metrics that are robust to skew, notably precision-recall curves and the area under the precision-recall curve.
\subsection{Preprocessing Pipeline}
\label{sec:preprocessing}
The dataset used throughout this study contains $1000$ accounts, of which $200$ ($20\%$) are labelled Suspicious and $800$ Benign, described by $33$ engineered behavioural features. This $80{:}20$ benign-to-suspicious ratio is considerably more balanced than the extreme skew observed in production AML systems (often well below $1\%$ suspicious). It was chosen deliberately to keep the minority class well represented at sample sizes that the quantum pipelines can process, while remaining skewed enough that accuracy alone is not meaningful: a rule that labels every account Benign is $80\%$ accurate and detects nothing. Precision, recall and F1 are therefore reported for the suspicious class throughout.
After feature generation, missing values arising mainly from accounts with no trading activity in a particular security category or window were imputed in a type-aware manner.
Ratio and temporal features, such as burstiness coefficients or the Hopkins statistic, were filled with the column-wise median computed on the training data.
Z-score features, which naturally centre at zero under neutral trading, were imputed to 0.0, and amount-based features (including co-trading metrics) were set to zero, correctly capturing the absence of the corresponding trading behaviour.
Following imputation, all features were standardised by removing the training-set mean and scaling to unit variance, so that features operating on different scales do not dominate distance-based or gradient-based learners.

The $33$ standardised features were then reduced to $20$ by one of two alternative methods, producing two fixed representations of the same accounts on which every model is evaluated:
\begin{itemize}
    \item \textbf{ANOVA}: univariate feature selection, retaining the $20$ original features with the largest one-way ANOVA $F$-statistic against the class label;
    \item \textbf{PCA}: projection onto the first $20$ principal components.
\end{itemize}
A third, non-linear $20$-dimensional embedding by uniform manifold approximation and projection (UMAP)~\cite{mcinnes2018umap} was explored in preliminary runs but did not outperform either representation for any model and is not reported.
Each reduction was fitted on the training partition and applied unchanged to the test partition, after which every component was rescaled to $[0,\pi]$ on the training data so that it can be used directly as a rotation angle in the quantum encodings of Sec.~\ref{sec:methods}. Because this rescaling is fitted on training data only, a few test values fall slightly outside the interval (ANOVA $-0.66$ to $3.51$; PCA $-0.15$ to $3.40$); they were neither clipped nor otherwise altered. Of the two, only ANOVA keeps original, interpretable features rather than mixtures of them.
\subsection{Train / Test Split Protocol}
All accounts were partitioned into a training set of $800$ accounts ($640$ benign, $160$ suspicious) and a held-out test set of $200$ accounts ($160$ benign, $40$ suspicious) by an $80{:}20$ stratified random split, preserving the benign-to-suspicious ratio in each subset. The split was performed purely at the account level, with no regard for which pump-and-dump event an account belonged to; this reflects a scenario in which the model must generalise to entirely new entities whose coordination patterns are unseen. A fixed random seed was used, so the assignment of individual accounts to train or test is deterministic and fully reproducible, and the same $200$ test accounts are used for every model and representation in Sec.~\ref{sec:results}. With $40$ suspicious test accounts, one additional detected account changes recall by $2.5$ percentage points, and one additional false alarm changes the false-positive rate by $0.625$ points; differences of a few points between models should be read with this granularity in mind.
% =====================================================================
\section{Methods}
\label{sec:methods}
\subsection{Classical Baselines}
\label{sec:methods:classical}
Four classical classifiers serve as reference points for the quantum models: L2-regularised logistic regression; gradient-boosted decision trees (XGBoost)~\cite{chen2016}; and two feed-forward neural networks with ReLU activations, dropout and a single logit output, NN1 with one hidden layer and NN3 with three. Each is trained and evaluated separately on the ANOVA and PCA representations of Sec.~\ref{sec:preprocessing}, giving eight configurations. Logistic regression and both networks receive an additional standard scaling fitted on the fitting portion of each cross-validation fold; XGBoost uses the representation directly. No resampling, additional features or synthetic observations are introduced.

Hyperparameters are selected by five-fold stratified cross-validation within the $800$ training accounts (seed $42$; $640$ fitting and $160$ validation accounts per fold). The search spaces are fixed in advance: for logistic regression, $C\in\{10^{-3},\dots,10^{2}\}$ with or without balanced class weights ($12$ candidates); for XGBoost, $20$ candidates over the number of trees ($100$, $300$, $600$), depth ($2$, $3$, $5$), learning rate, child weight, row and column subsampling, L2 penalty and class weighting; for NN1 (hidden width $16$, $32$ or $64$) and NN3 (widths $32$--$16$--$8$ or $64$--$32$--$16$), $20$ candidates each over learning rate, weight decay, dropout, epoch budget and class weighting. The networks are trained with Adam~\cite{kingma2015adam} on binary cross-entropy with batch size $64$ and no early stopping, so that the validation folds are never used to decide when to stop.

Every candidate produces one out-of-fold score per training account. The decision threshold is chosen by sweeping every distinct out-of-fold score and selecting the one that maximises pooled out-of-fold F1 for the suspicious class; the candidate and threshold with the highest such F1 are then frozen, the model is refitted on all $800$ training accounts, and it is scored once on the $200$ test accounts. No test outcome is used to choose a model, a hyperparameter or a threshold. As a secondary comparison, the same test scores are also evaluated at the default threshold of $0.5$. XGBoost, NN1 and NN3 are additionally refitted under nine further random seeds with their selected hyperparameters held fixed, each seed obtaining its own training-side threshold, to measure sensitivity to training randomness.

For the QKNN specifically, a classical $k$-nearest-neighbours classifier ($k=5$, Minkowski distance, uniform voting) on the same split and representations serves as the like-for-like control.
\subsection{Quantum $k$ Nearest Neighbors (QKNN)}
\label{sec:methods:qknn}
Unlike conventional machine learning algorithms that explicitly learn decision boundaries, QKNN is an instance-based classifier that predicts the class of an unseen sample by measuring its similarity to previously observed training instances.
\subsubsection{Standard QKNN}
The standard QKNN performs classification by assigning the label of a test sample based on the majority class among its nearest neighbours in the training dataset. Given a classical dataset,
\begin{equation}
\mathcal{D}=\{(\mathbf{x}_i,y_i)\}_{i=1}^{N},
\end{equation}
where $\mathbf{x}_i\in\mathbb{R}^{d}$ denotes the $d$-dimensional feature vector and $y_i$ represents the corresponding class label, each feature vector is first normalized and encoded into a quantum state using a quantum feature map,
\begin{equation}
\mathbf{x}_i
\longrightarrow
|\phi(\mathbf{x}_i)\rangle.
\end{equation}
The similarity between a test sample $\mathbf{x}$ and a training sample $\mathbf{x}_i$ is quantified using the quantum state fidelity,
\begin{equation}
F(\mathbf{x},\mathbf{x}_i)
=
\left|
\left\langle
\phi(\mathbf{x})
\middle|
\phi(\mathbf{x}_i)
\right\rangle
\right|^{2},
\end{equation}
where larger fidelity values indicate greater similarity between the encoded quantum states. The fidelity values are estimated using the Compute--Uncompute algorithm, which computes the overlap between two quantum states by preparing one quantum state, applying the inverse circuit of the second state, and measuring the probability of obtaining the all-zero computational basis state. This method enables efficient fidelity estimation while remaining compatible with current noisy intermediate-scale quantum (NISQ) hardware.
After computing the fidelity values between the test sample and every training sample, the samples are ranked according to their similarity,
\begin{equation}
F_{1}\geq F_{2}\geq\cdots\geq F_{N},
\end{equation}
and the top-$k$ nearest neighbours are selected,
\begin{equation}
\mathcal{N}_{k}(\mathbf{x})
=
\left\{
x_{(1)},x_{(2)},\ldots,x_{(k)}
\right\}.
\end{equation}
The predicted class label is obtained using majority voting,
\begin{equation}
\hat{y}
=
\arg\max_{c}
\sum_{i\in\mathcal{N}_{k}(\mathbf{x})}
\mathbb{I}(y_i=c),
\end{equation}
where $\mathbb{I}(\cdot)$ denotes the indicator function. The workflow of the standard QKNN classifier is summarized as follows:
\begin{enumerate}
    \item Normalize the classical feature vectors.
    \item Encode each sample into a quantum state using a quantum feature map.
    \item Compute quantum state fidelities between the test sample and all training samples using the Compute--Uncompute algorithm.
    \item Rank the training samples according to their fidelity scores.
    \item Select the top-$k$ nearest neighbours.
    \item Predict the class label using majority voting.
\end{enumerate}
\subsubsection{Batch Processing Variant}
The computational cost of QKNN is dominated by repeated fidelity evaluations between each test sample and every training sample. A straightforward implementation requires constructing and executing one quantum circuit for every pair of test and training samples, leading to significant circuit compilation and execution overhead.
To improve hardware utilization and reduce execution latency, we propose a hardware-aware batching strategy that exploits the available qubit capacity of the target quantum processor. Instead of evaluating a single training sample per circuit execution, multiple independent fidelity computations are packed into a single reusable quantum circuit. Each fidelity computation occupies an independent region of the quantum register, allowing several train samples to be evaluated simultaneously during one circuit execution.
Unlike algorithmic modifications to QKNN, the proposed batching strategy does not alter the similarity measure or the classification rule. It is purely an execution-level optimization that maximizes hardware utilization while preserving identical classification results.

\paragraph{Batch Size Estimation.}
For Dense Angle Encoding, a single data point requires
\begin{equation}
q
=
\left\lceil
\frac{N}{2}
\right\rceil
\end{equation}
qubits for state preparation, where \(N\) denotes the number of encoded features.
Given a quantum device containing \(H\) available qubits, the maximum number of independent fidelity computations that can be executed within a single circuit is
\begin{equation}
B
=
\left\lfloor
\frac{H}{q}
\right\rfloor .
\end{equation}
Consequently, each circuit execution processes at most \(B\) training samples simultaneously while fully utilizing the available hardware resources.

\paragraph{Sequential Batch Execution.}
For a given test sample \(x_{\mathrm{test}}\), the training dataset is partitioned into batches,
\begin{equation}
X_{\mathrm{train}}^{(1)},
X_{\mathrm{train}}^{(2)},
\dots,
X_{\mathrm{train}}^{(m)},
\end{equation}
where each batch contains at most \(B\) training samples.
During execution, the same test sample is broadcast to every computation slot, while each slot encodes a different training sample. Since every slot occupies an independent subset of qubits, the fidelity computations are mutually independent and can be evaluated simultaneously within a single circuit execution.
For one batch, the circuit therefore computes
\begin{equation}
\left[
K_1,
K_2,
\dots,
K_B
\right],
\end{equation}
where
\begin{equation}
K_i
=
\left|
\left\langle
\psi(x_i)
\middle|
\psi(x_{\mathrm{test}})
\right\rangle
\right|^2.
\end{equation}
After completing one batch, the next batch is loaded into the same circuit until fidelities with all training samples have been evaluated.

\paragraph{Kernel Matrix Construction.}
The reusable circuit is executed sequentially over all training batches for each test sample. The resulting fidelity values form one complete row of the quantum kernel matrix,
\begin{equation}
\mathbf{K}_{\mathrm{test}}
=
\left[
K(x_{\mathrm{test}},x_1),
K(x_{\mathrm{test}},x_2),
\ldots,
K(x_{\mathrm{test}},x_M)
\right].
\end{equation}
Repeating the same procedure for every test sample produces the complete kernel matrix,
\begin{equation}
\mathbf{K}
\in
\mathbb{R}^{M_{\mathrm{test}}\times M_{\mathrm{train}}}.
\end{equation}

\paragraph{Reusable Circuit Architecture.}
The quantum circuit is constructed and transpiled only once. Throughout execution, the circuit topology remains unchanged, including:
\begin{itemize}
    \item Qubit allocation,
    \item Register organization,
    \item Computation-slot layout,
    \item Transpilation,
    \item Hardware routing.
\end{itemize}
Only the circuit parameters corresponding to the test sample and the current training batch are updated through parameter binding before each execution. This eliminates repeated circuit construction and transpilation overhead while enabling efficient reuse of the same executable quantum circuit.

\paragraph{Computational Complexity.}
For a training dataset containing \(M\) samples and a test dataset containing \(M_{\mathrm{test}}\) samples, the conventional QKNN implementation requires
\begin{equation}
M_{\mathrm{test}}
\times
M
\end{equation}
independent circuit executions.
Using the proposed batching strategy, the number of circuit executions becomes
\begin{equation}
M_{\mathrm{test}}
\times
\left\lceil
\frac{M}{B}
\right\rceil
\end{equation}
Although the total number of fidelity evaluations remains unchanged, batching substantially reduces circuit execution overhead by decreasing the number of circuit launches, minimizing repeated transpilation, reducing backend scheduling overhead, and maximizing utilization of the available quantum hardware.
\subsubsection{Voting Strategies}

\paragraph{Weighted Fidelity Voting.}
In this formulation, all neighbors contribute equally irrespective of their fidelity similarity magnitude. To incorporate similarity-sensitive voting, fidelity weighted voting assigns larger voting contribution to data points with stronger quantum similarity.
For a given test data point $x_t$, the first-order quantum neighborhood is defined as:
\begin{equation}
\mathcal N^{(1)}(x_t)
=
\operatorname{TopK}
\{
K(x_t,x_i)
\}_{i=1}^{M}.
\end{equation}
The fidelity weighted voting score for class $c$ is computed as:
\begin{equation}
V_c^{(1)}(x_t)
=
\sum_{x_i\in\mathcal N^{(1)}(x_t)}
K(x_t,x_i)
\mathbf 1(y_i=c).
\end{equation}
The predicted label becomes:
\begin{equation}
\hat y
=
\arg\max_c
V_c^{(1)}(x_t).
\end{equation}
In this formulation:
\begin{itemize}
    \item Neighbors with larger fidelity similarity contribute stronger voting influence,
    \item Weakly similar neighbors contribute proportionally less,
    \item Local fidelity geometry directly influences prediction.
\end{itemize}

\paragraph{Mean Weighted Fidelity Voting.}
For highly imbalanced datasets, fidelity weighted voting may still become biased toward majority classes when one class contains more neighbors. To stabilize the voting contribution across classes, mean fidelity weighted voting computes the average fidelity contribution for each class. For a given class $c$, the class-wise mean fidelity score becomes:
\begin{equation}
S_c(x_t)
=
\frac{
\sum_{x_i\in\mathcal N^{(1)}(x_t)}
K(x_t,x_i)
\mathbf 1(y_i=c)
}{
\sum_{x_i\in\mathcal N^{(1)}(x_t)}
\mathbf 1(y_i=c)
}.
\end{equation}
The predicted label becomes:
\begin{equation}
\hat y
=
\arg\max_c
S_c(x_t).
\end{equation}
The voting then depends on the average fidelity similarity within each class. Minority classes are not suppressed purely due to lower neighbor count, and strongly similar suspicious neighbors can dominate weakly similar majority neighbors.
\paragraph{Anomaly Amplified Voting.}
In anomaly detection problems, suspicious samples are often sparse and distributed across isolated regions of the feature space, limiting their influence during majority voting.
To strengthen the representation of anomalous regions, we introduce an anomaly-amplified voting strategy in which confidently predicted suspicious samples are recursively incorporated into the voting memory. The intuition is that newly identified suspicious samples can reinforce local neighbourhoods and increase the likelihood of correctly identifying subsequent anomalies that exhibit similar quantum representations.
Let the voting memory at inference step \(t\) be defined as
\begin{equation}
\mathcal{D}^{(t)}
=
\left\{
(x_i,y_i)
\right\}_{i=1}^{M_t},
\end{equation}
where \(M_t\) denotes the number of samples currently stored in the voting memory. For a test sample \(x_t\), the nearest neighbours are determined by computing the quantum fidelity between \(x_t\) and all samples contained in \(\mathcal{D}^{(t)}\), followed by fidelity-weighted majority voting.
After the prediction is obtained, the voting memory is updated according to the predicted class label. If the predicted label corresponds to a suspicious transaction, $\hat{y}_t=-1$, the sample is incorporated into the voting memory,
\begin{equation}
\mathcal{D}^{(t+1)}
=
\mathcal{D}^{(t)}
\cup
\left\{
(x_t,-1)
\right\}.
\end{equation}
Otherwise, if the predicted label corresponds to a benign transaction, $\hat{y}_t=+1$, the voting memory remains unchanged, $\mathcal{D}^{(t+1)} = \mathcal{D}^{(t)}$.
Consequently, the neighborhood used for future predictions evolves dynamically during inference. For the subsequent test sample \(x_{t+1}\), the neighborhood is computed using the updated voting memory,
\begin{equation}
\mathcal{N}_k(x_{t+1})
=
\operatorname{TopK}
\left\{
K(x_{t+1},x_i)
\right\}_{x_i\in\mathcal{D}^{(t+1)}}.
\end{equation}
As additional suspicious samples are accumulated, the voting memory progressively expands the representation of anomalous regions within the quantum similarity space. This results in an evolving inference process,
\begin{equation}
\mathcal{D}^{(1)}
\subseteq
\mathcal{D}^{(2)}
\subseteq
\cdots
\subseteq
\mathcal{D}^{(T)},
\end{equation}
where the influence of suspicious neighbourhoods increases over successive prediction steps.
Rather than relying on a fixed reference dataset, the classifier progressively strengthens the representation of sparse anomalous regions, enabling future test samples with similar quantum fidelity characteristics to receive greater voting support.
\subsubsection{Higher-Order Quantum k-Nearest Neighbour}
The standard QKNN classifier retrieves the top-$k$ neighbours of a test sample using only the first-order quantum fidelity similarities. For a test sample $x_t$, the first-order neighbourhood is defined as
\begin{equation}
\mathcal N^{(1)}(x_t)
=
\operatorname{TopK}
\left\{
K(x_t,x_i)
\right\}_{i=1}^{M},
\end{equation}
where $K(x_t,x_i)$ denotes the quantum fidelity between the test sample and the $i$-th training sample.
The class prediction is obtained through mean-fidelity weighted voting,
\begin{equation}
V_c^{(1)}(x_t)
=
\frac{
\sum\limits_{x_i\in\mathcal N^{(1)}(x_t)}
K(x_t,x_i)\,
\mathbb I(y_i=c)
}{
\sum\limits_{x_i\in\mathcal N^{(1)}(x_t)}
\mathbb I(y_i=c)
},
\end{equation}
and
\begin{equation}
\hat y
=
\arg\max_c
V_c^{(1)}(x_t).
\end{equation}
Although effective, the first-order neighbourhood considers only direct fidelity relationships between the test sample and individual training samples. It ignores indirect neighbourhood information that may be revealed through intermediate samples within the fidelity graph.
To capture these indirect relationships, we introduce higher-order quantum neighbourhood propagation. Instead of relying solely on the original fidelity matrix $K \in \mathbb R^{M\times M}$, the neighbourhood information is recursively propagated through successive matrix powers. The $r$-th order propagated fidelity matrix is given by $K^{(r)} = K^r$, where $K^2, K^3, \dots, K^R$ capture two-hop, three-hop, and higher-order quantum neighbourhood relationships, respectively.
The overall propagated similarity matrix is then constructed as
\begin{equation}
\widetilde K
=
\sum_{r=1}^{R}
\alpha_r
K^r,
\end{equation}
where $\alpha_r$ controls the contribution of the $r$-th propagation order.
To prevent distant neighbourhoods from dominating the similarity landscape, exponentially decaying propagation weights are employed,
\begin{equation}
\alpha_r
=
\lambda^{\,r-1},
\qquad
0<\lambda<1.
\end{equation}
The propagated kernel therefore incorporates both direct and indirect fidelity relationships while assigning progressively smaller importance to higher-order neighbourhoods.
Neighbour retrieval is subsequently performed using the propagated similarity matrix,
\begin{equation}
\widetilde{\mathcal N}^{(1)}(x_t)
=
\operatorname{TopK}
\left\{
\widetilde K(x_t,x_i)
\right\}_{i=1}^{M},
\end{equation}
and the final prediction is obtained through mean-fidelity weighted voting,
\begin{equation}
V_c(x_t)
=
\frac{
\sum\limits_{x_i\in
\widetilde{\mathcal N}^{(1)}(x_t)}
\widetilde K(x_t,x_i)
\mathbb I(y_i=c)
}{
\sum\limits_{x_i\in
\widetilde{\mathcal N}^{(1)}(x_t)}
\mathbb I(y_i=c)
},
\end{equation}
followed by $\hat y = \arg\max_c V_c(x_t)$.
The proposed higher-order QKNN therefore extends conventional neighbour retrieval from direct quantum fidelity to a multi-scale propagated similarity representation, allowing information to diffuse through the quantum neighbourhood graph before classification.
\subsubsection{Fidelity-Based Clustering QKNN}
As there is no training phase in QKNN, the label assigned to a given test point $x_t$ is determined entirely by the voting scheme applied at inference time. This motivates partitioning the training data into fidelity-based clusters, each with a cluster-level representative state; for a given test point $x_t$, the fidelity score against these representative states is computed, and the maximum of these scores identifies the most likely cluster, within which a local QKNN determines the label.

\paragraph{Defining the closeness of two points.} Unlike the classical regime, in the quantum regime we have fidelity. For two points $x_1$, $x_2$, the fidelity score gives the similarity of the two points. The distance definition can then be

\begin{equation}
D_{ij} = 1 - F_{ij}.
\end{equation}

\paragraph{Intra-Cluster Distance.}
The intra-cluster distance is computed as the average pairwise distance within the cluster:
\begin{equation}
D_{\mathrm{intra}}(C_r)
=
\frac{1}{|C_r|^2}
\sum_{i,j \in C_r}
D_{ij}.
\end{equation}
The representative state is the medoid of the cluster, identified during the clustering process and used to compute distances between clusters.
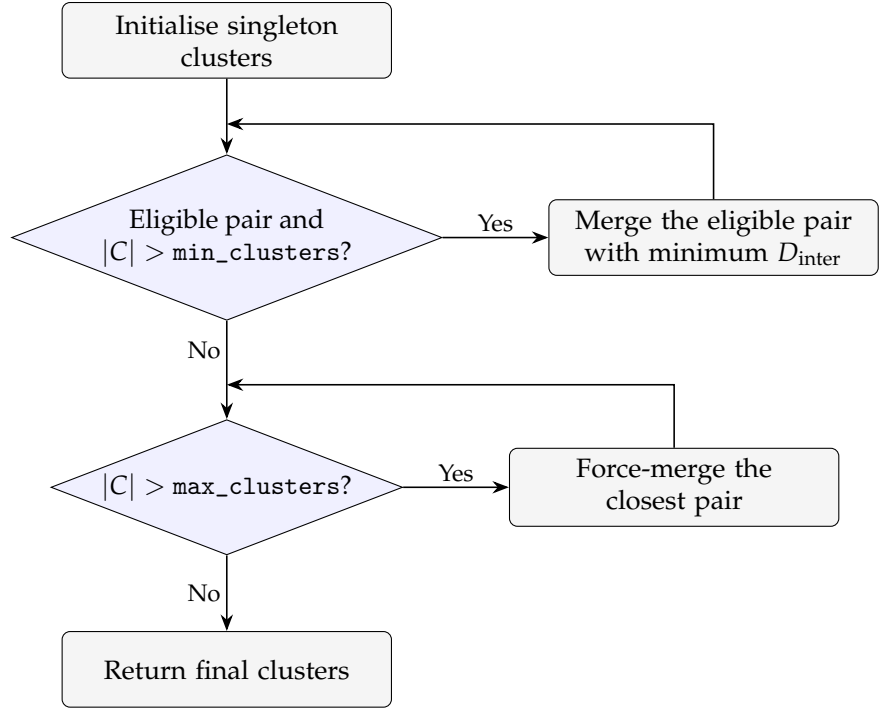
\begin{figure}[H]
    \centering
    \begin{tikzpicture}[
        node distance=10mm and 14mm,
        >={Stealth[length=2.2mm]},
        box/.style={draw, rounded corners=3pt, fill=gray!8, align=center,
                    text width=41mm, minimum height=10mm, font=\small},
        dec/.style={draw, diamond, aspect=2.6, fill=blue!6, align=center,
                    inner sep=1.5pt, font=\small},
        lab/.style={font=\footnotesize, inner sep=2pt},
        flow/.style={->, semithick}
    ]
    \node[box] (init) {Initialise singleton\\ clusters};
    \node[dec, below=of init] (elig) {Eligible pair and\\ $|C| > \texttt{min\_clusters}$?};
    \node[box, right=of elig] (merge) {Merge the eligible pair\\ with minimum $D_{\mathrm{inter}}$};
    \node[dec, below=13mm of elig] (maxc) {$|C| > \texttt{max\_clusters}$?};
    \node[box, right=of maxc] (force) {Force-merge the\\ closest pair};
    \node[box, below=10mm of maxc] (done) {Return final clusters};

    \draw[flow] (init) -- (elig);
    \draw[flow] (elig) -- node[lab, above] {Yes} (merge);
    \draw[flow] (merge.north) |- ($(init.south)!0.6!(elig.north)$);
    \draw[flow] (elig) -- node[lab, left, pos=0.3] {No} (maxc);
    \draw[flow] (maxc) -- node[lab, above] {Yes} (force);
    \draw[flow] (force.north) |- ($(elig.south)!0.65!(maxc.north)$);
    \draw[flow] (maxc) -- node[lab, left] {No} (done);
    \end{tikzpicture}
    \caption{Fidelity-aware agglomerative clustering workflow. Eligible pairs are merged in order of increasing medoid distance until no eligible pair remains or the minimum number of clusters is reached; if more than the maximum number of clusters remain, the closest pairs are then force-merged.}
    \label{fig:fidelity_clustering}
\end{figure}

\paragraph{Inter-Cluster Distance.}
For two clusters \(C_r\) and \(C_s\), the inter-cluster distance is computed using the distance between their medoids:
\begin{equation}
D_{\mathrm{inter}}(C_r,C_s)
=
D\!\left(
\mathrm{medoid}(C_r),
\mathrm{medoid}(C_s)
\right).
\end{equation}

\paragraph{Agglomerative Merging.}
Clusters are merged iteratively using a Ward-style criterion. A merge between clusters \(C_r\) and \(C_s\) is permitted when $D_{\mathrm{inter}}(C_r,C_s) \le \tau$ and $D_{\mathrm{inter}}(C_r,C_s) < D_{\mathrm{intra}}(C_r) + D_{\mathrm{intra}}(C_s)$, where \(\tau\) is the inter-cluster distance threshold. Among all eligible pairs, the pair with the smallest inter-cluster distance is merged first.

\paragraph{Test Sample Routing.}
For a test sample \(x_t\), the fidelity similarity vector to the training set is computed. The routing distance to cluster \(C_r\) is given by
\begin{equation}
D(x_t,C_r)
=
1 -
F\!\left(
x_t,
\mathrm{medoid}(C_r)
\right),
\end{equation}
and the test sample is assigned to the nearest cluster $C^* = \arg\min_r D(x_t,C_r)$.

\paragraph{Local QKNN Classification.}
After routing, neighbour selection is restricted to the members of the assigned cluster \(C^*\). The top-\(k\) neighbours are selected using fidelity similarity and classification is performed using mean-fidelity weighted voting.
The overall prediction pipeline is therefore
\begin{equation}
\begin{gathered}
x_t \;\rightarrow\; \text{Cluster Routing} \;\rightarrow\; \text{Local Top-}k \\
\;\rightarrow\; \text{Mean-Fidelity Voting} \;\rightarrow\; \hat{y}.
\end{gathered}
\end{equation}
\subsection{Related Work on QKNN Voting and Clustering Strategies}
\label{sec:qknn-related-work}
\subsubsection{Quantum k-Nearest Neighbour and Voting Strategies}
The k-Nearest Neighbour (KNN) classifier is one of the earliest and most widely adopted instance-based learning algorithms. In its original formulation, the class label of a query sample is determined through majority voting among its $k$ nearest neighbours, where each neighbour contributes equally irrespective of its distance from the query point. Although simple and effective, equal voting often results in poor robustness in the presence of noisy observations, class imbalance, and heterogeneous data distributions. Consequently, numerous variants of KNN have focused on improving the voting mechanism while preserving the underlying neighbour retrieval strategy.
The earliest extension was the inverse-distance weighted KNN proposed by Dudani \cite{dudani1976}, in which neighbours closer to the query sample contribute proportionally larger voting weights than distant neighbours. Alternative weighting functions, including Gaussian kernel weighting \cite{hechenbichler2004weighted}, adaptive weighting \cite{lu2008adaptive}, and dual-weighted KNN that jointly considers neighbour distance and ranking \cite{gou2012dual}, have subsequently been proposed to further improve local decision boundaries. Collectively, these approaches demonstrate that assigning weights according to neighbour reliability consistently improves classification performance over conventional majority voting, particularly in noisy and imbalanced datasets.
Another important direction is \emph{Fuzzy KNN}, introduced by Keller \emph{et al.} \cite{keller1985}, in which neighbours contribute fuzzy membership values instead of discrete class labels. Rather than assigning equal class votes, each neighbour possesses a degree of membership to every class, allowing the final prediction to be obtained through weighted aggregation of these memberships.
More recently, several studies have explored reliability-aware voting mechanisms that seek to reduce the influence of noisy or uninformative neighbours. Representative approaches include dual-weighted KNN, which jointly incorporates neighbour distance and rank into the voting weight \cite{gou2012dual}; mutual k-nearest neighbour methods, where a neighbour contributes only if the neighbourhood relationship is reciprocal, thereby improving robustness against isolated samples \cite{ozaki2011mutual}; certainty-based voting schemes that assign weights according to the local confidence or certainty of each neighbour \cite{wang2007local}; and evolutionary or adaptive voting strategies that optimize neighbour weights using evolutionary algorithms or meta-heuristic optimization techniques \cite{garcia2012prototype}. Collectively, these methods demonstrate that incorporating neighbourhood reliability into the voting process substantially improves classification performance without modifying the underlying neighbour search algorithm.
The Quantum k-Nearest Neighbour (QKNN) naturally inherited these concepts from classical KNN. Existing QKNN algorithms employ quantum state fidelity as the similarity measure, where the overlap between quantum feature states replaces classical distance computations. Basheer \emph{et al.} \cite{basheer2020qknn} proposed one of the first practical fidelity-based Quantum k-Nearest Neighbour algorithms, where neighbour retrieval is performed directly using quantum state fidelity. Their formulation demonstrated that quantum fidelity can effectively replace classical distance metrics while preserving the non-parametric nature of KNN classification. Subsequent research has focused on improving the efficiency of QKNN through divide-and-conquer neighbour search \cite{gong2024divide}, alternative quantum distance formulations including Euclidean and Mahalanobis distance estimation \cite{zardini2024euclidean,gao2022mahalanobis}, and hardware-oriented implementations suitable for noisy intermediate-scale quantum (NISQ) devices.
Despite these advances, the prediction stage of QKNN has remained largely unchanged. Existing algorithms typically compute the fidelity between a query sample and every training sample, retrieve the top-$k$ neighbours according to the fidelity ranking, and assign the final class label using majority voting or direct fidelity ranking. Consequently, neighbour retrieval is entirely based on first-order pairwise similarity, while each prediction is performed independently without exploiting structural relationships between neighbouring samples.
Compared with the extensive literature on voting strategies in classical KNN, relatively little attention has been devoted to the voting mechanism in QKNN. Most existing works employ majority voting after neighbour retrieval, while a limited number of studies utilize fidelity scores solely for neighbour ranking rather than as explicit voting weights. Furthermore, existing QKNN methods generally treat each neighbour independently and do not incorporate higher-order neighbourhood information, adaptive voting memories, or locality-aware routing strategies during inference.
In this work, we adopt a \emph{mean fidelity-aware weighted voting} strategy, in which each neighbour contributes proportionally to its quantum fidelity with the query sample. Since quantum fidelity directly quantifies the overlap between two quantum feature states, it provides a natural measure of neighbour reliability. Consequently, neighbours exhibiting higher fidelity exert greater influence on the final prediction, while weakly related neighbours contribute proportionally less. Building upon this fidelity-aware voting formulation, we further introduce \textit{higher-order fidelity propagation}, \textit{anomaly-aware dynamic voting}, and \textit{fidelity-based clustering} to exploit both direct and indirect quantum neighbourhood structures while remaining fully compatible with existing quantum kernel estimation techniques.
\subsubsection{Higher-Order Fidelity Propagation}
Higher-order neighbourhood learning has been extensively studied in graph-based machine learning as an effective means of capturing structural relationships beyond immediate neighbours. Early work by Zhou \emph{et al.} introduced label propagation through local and global consistency, demonstrating that iteratively propagating information across neighbouring nodes significantly improves classification performance by exploiting the underlying manifold structure of the data \cite{zhou2004learning}. Building upon this idea, graph neural network architectures, including Graph Convolutional Networks (GCN) \cite{kipf2017semi} and Approximate Personalized Propagation of Neural Predictions (APPNP) \cite{klicpera2019predict}, aggregate information from multi-hop neighbourhoods to learn more expressive node representations. Similarly, diffusion kernels \cite{kondor2002diffusion} and iterative higher-order propagation methods \cite{mahesh2022ihop} model indirect relationships through repeated propagation over graph structures, while recent higher-order diffusion networks have shown that information from higher-order neighbours can preserve discriminative label information even when first-order neighbourhoods are sparse or noisy \cite{li2024hidnet}.
Although higher-order neighbourhood propagation has been widely explored in graph learning and network analysis, its application to Quantum k-Nearest Neighbour (QKNN) classification remains largely unexplored. Existing QKNN algorithms perform neighbour retrieval exclusively using first-order quantum fidelity relationships, where each test sample is compared independently with the training samples based solely on pairwise quantum state similarity. Consequently, indirect relationships between neighbouring samples are not incorporated into the prediction process.
Inspired by higher-order neighbourhood propagation in graph learning, this work extends QKNN by introducing higher-order propagation directly on the quantum fidelity matrix. Rather than performing neighbour retrieval using only the first-order fidelity scores, the proposed framework propagates quantum similarity across multiple neighbourhood orders to construct a richer fidelity representation. This enables indirect quantum neighbourhood relationships to contribute to neighbour retrieval while preserving the underlying quantum kernel and maintaining compatibility with existing fidelity-based QKNN frameworks.
\subsubsection{Fidelity-Based Clustering Algorithm}
\label{sec:fidelity_based_clustering}
Clustering has been investigated as a means of exploiting local similarity and structural relationships within data. In quantum machine learning, recent work has extended this idea by using quantum-state similarity measures to construct clustering structures. In particular, Xiong \emph{et al.} proposed QGKM, a quantum fidelity-based graph clustering framework that maps classical samples into quantum states and uses quantum fidelity to construct a similarity structure. The framework further incorporates a $k$-nearest-neighbour graph to preserve local geometric relationships and employs a deterministic greedy hierarchical merging strategy for cluster formation \cite{xiong2026qgkm}.
QGKM demonstrates that quantum fidelity can provide a meaningful similarity representation for clustering, particularly for data exhibiting nonlinear, heterogeneous, and complex structures. However, the clustering objective in QGKM is primarily unsupervised: the fidelity-based graph and subsequent merging procedure are used to obtain the final clusters, which are not subsequently incorporated into a quantum nearest-neighbour classification procedure.
This distinction is important for the present work. Rather than treating fidelity-based clustering as the final learning task, we investigate its use as an intermediate structure for QKNN inference. The proposed approach uses the fidelity-induced similarity space to partition the training data and subsequently routes a test sample to a locally relevant cluster before performing QKNN classification. Thus, the role of clustering is changed from final pattern discovery to neighbourhood localization for supervised classification. The resulting framework therefore connects two lines of research that have largely been investigated independently, quantum-fidelity-based clustering and fidelity-based QKNN, using clustering to exploit the local structure of the quantum fidelity landscape while retaining the original QKNN classification mechanism within the selected neighbourhood. The experimental configuration and results for this method are reported in Sec.~\ref{sec:qknn-results}.
\subsection{Quantum Neural Network (QNN) for Classification}
\label{sec:methods:hqnn}
Quantum neural networks (QNNs), also called variational quantum classifiers, are the supervised-learning members of the broader family of variational quantum algorithms \cite{cerezo2021vqa,bharti2022nisq}. The defining object is a parameterized quantum circuit (PQC) \cite{benedetti2019pqc} whose gate angles are split into two groups: those that carry the classical input $\mathbf{x}$ into the Hilbert space, and those that are free parameters $\boldsymbol{\theta}$ optimised against a training objective. Measurement of a fixed observable produces a real-valued score that is thresholded or squashed into a class probability. The template was introduced independently by Farhi and Neven \cite{farhi2018qnn} and by Mitarai \emph{et al.} \cite{mitarai2018qcl}.
\subsubsection{Architecture}
\begin{figure}[H]
    \centering
    \includegraphics[width=0.5\linewidth]{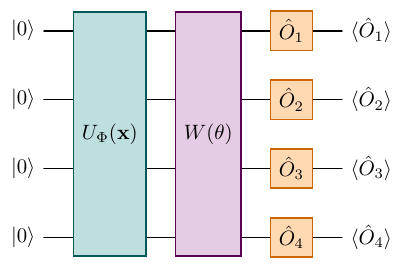}
    \caption{Schematic showing the basic QNN circuit. $U_{\Phi}(\mathbf{x})$ is the feature map, which maps the $n$-dimensional classical vector $\mathbf{x}$ to a $2^n$-dimensional quantum state vector. $W(\mathbf{\theta})$ is the parameterized circuit, which consists of trainable parameters $\mathbf{\theta}$ that are optimized over multiple iterations and many loss functions. Finally, we measure the expectation of a local observable $\hat{O}_i$, which in our case is the Pauli $Z$ operator.}
    \label{fig:qnn}
\end{figure}
A QNN consists of three components (see Fig.~\ref{fig:qnn}): (i) a feature map $S(\mathbf{x})$ to load the classical data into the quantum circuit, (ii) a variational circuit $W(\mathbf{\theta})$ which consists of trainable parameters $\mathbf{\theta}$, and (iii) an observable, whose expectation value is used to make a prediction. This is augmented by a classical feature-compression front end and a classical decision layer at the end. We describe each in turn, before introducing the hybrid extension of this architecture (HQNN, Sec.~\ref{sec:hqnn}), whose results are reported separately in Sec.~\ref{sec:hqnn-results}.
\subsubsection{Classical Feature Compression}
Raw surveillance features rolling returns, realized volatility, volume $z$-scores, order-book imbalance, trade-count bursts, and their lags live in a space of dimension far exceeding what can be encoded on a near-term device. Since angle-type encodings consume one rotation angle per feature per layer, the qubit budget $n$ directly bounds the input dimension $d$. The QNN therefore operates on one of the $20$-dimensional representations of Sec.~\ref{sec:preprocessing} (ANOVA selection or PCA), each already rescaled to $[0,\pi]$ so that encoding angles occupy the full period without aliasing, and encodes one feature per qubit on a $20$-qubit register.
\subsubsection{Quantum Feature Maps}
The reduced vector $\mathbf{x}\in\mathbb{R}^{d}$ is embedded via
\begin{align}
|\Phi(\mathbf{x})\rangle &= U_\Phi(\mathbf{x})\,H^{\otimes n}\,|0\rangle^{\otimes n}, \\
U_\Phi(\mathbf{x}) &= \exp\!\Big(i\sum_{S\subseteq[n]}\phi_S(\mathbf{x})\prod_{i\in S}Z_i\Big),
\end{align}
restricted to $|S|\le 2$, which is the second-order ZZ feature map of \cite{havlicek2019supervised} with $\phi_{\{i\}}(\mathbf{x})=x_i$ and $\phi_{\{i,j\}}(\mathbf{x})=(x_i \, x_j) / \pi$ \cite{jha2026comparative}. The diagonal-in-$Z$ structure means the map is realised by single-qubit $R_Z$ rotations and CNOT-sandwiched $R_Z$ pairs, keeping two-qubit depth linear in the number of entangling pairs. The pairwise terms are what give the map non-trivial correlation structure: with $|S|\le1$ the induced kernel factorises into a product of single-qubit kernels and is classically simulable, whereas the second-order map is believed to evade efficient classical estimation \cite{havlicek2019supervised}. We treat the entangling pattern (linear versus full) as a hyperparameter, noting that denser patterns increase expressibility and simultaneously worsen kernel concentration \cite{thanasilp2024concentration,holmes2022expressibility}.
\subsubsection{Hybrid Extension: Hybrid Quantum Neural Network (HQNN)}
\label{sec:hqnn}
\begin{figure}[H]
\centering
\includegraphics[width=\textwidth]{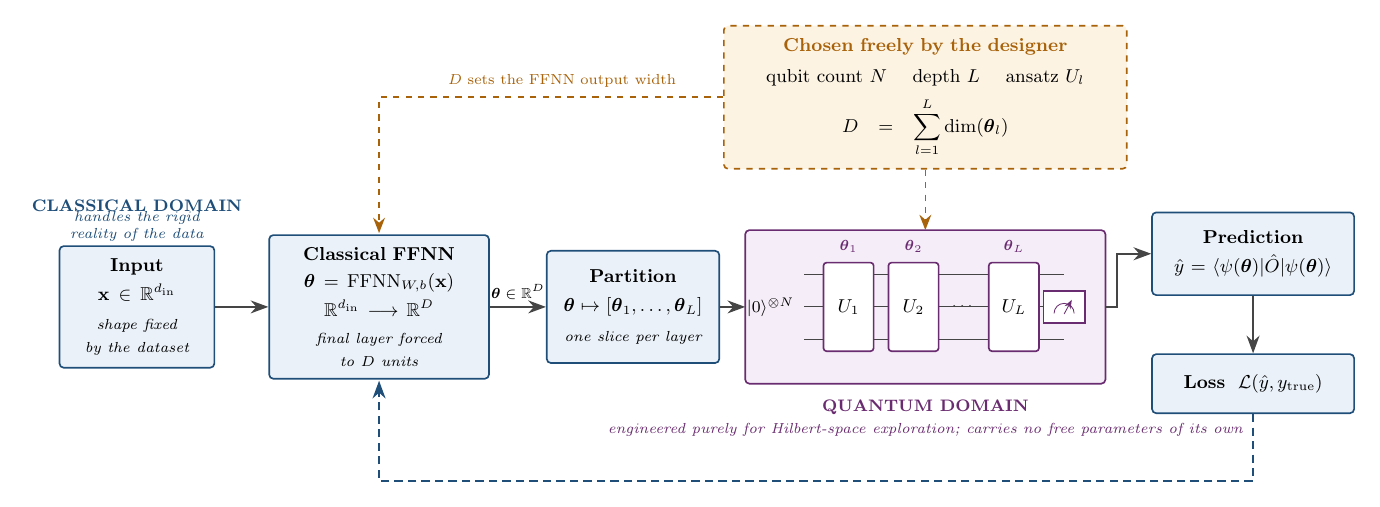}
\caption{\label{fig:hqnn} Hybrid architecture with generalized data re-uploading. The FFNN absorbs the input dimension $d_{\mathrm{in}}$ and emits exactly the $D$ angles the circuit requires, so $(N,L)$ may be varied without altering the data pipeline. Dashed paths show the circuit choices fixing the classical output width (forward) and gradient flow to the only trainable parameters, $(W,b)$ (return).}
\end{figure}

In the conventional variational classifier described above, the input is written into the circuit through a fixed feature map, so the register width is dictated by the data dimension: $20$ features require $20$ qubits. The hybrid quantum neural network (HQNN) removes this coupling. It is a generalized data re-uploading architecture that interposes a classical feed-forward network (FFNN) which \emph{generates} the circuit parameters rather than letting the circuit consume the features directly, so that every layer's rotation angles are a learned function of the input rather than fixed once by a static encoding gate. The model is the composition
\begin{equation}
\label{eq:hqnn-map}
\begin{split}
\mathbf{x} &\;\longmapsto\; \boldsymbol{\theta} = \mathrm{FFNN}_{W,b}(\mathbf{x}) \\
&\;\Longrightarrow\; |\psi(\boldsymbol{\theta})\rangle = \prod_{l=1}^{L} U_l(\boldsymbol{\theta}_l)\,|0\rangle^{\otimes N} \\
&\;\Longrightarrow\; \hat{y} = \langle\psi(\boldsymbol{\theta})|\hat{O}|\psi(\boldsymbol{\theta})\rangle ,
\end{split}
\end{equation}
with $\boldsymbol{\theta}=[\boldsymbol{\theta}_1,\dots,\boldsymbol{\theta}_L]$, as shown in Fig.~\ref{fig:hqnn}.
The construction is ordered: the circuit is specified first, and the classical network is sized to match. Given $N$, $L$ and the generator $U_l$, the quantum parameter dimension is $D = \sum_{l=1}^{L}\dim(\boldsymbol{\theta}_l)$. The terminal FFNN layer is constrained to emit exactly $D$ units, $\mathrm{FFNN}_{W,b}:\mathbb{R}^{d_{\mathrm{in}}}\!\to\!\mathbb{R}^{D}$, and the output is sliced into contiguous sub-vectors, one per layer. Since only the terminal width changes, the hidden architecture and the entire preprocessing chain are untouched when the circuit is altered: a two-qubit and a five-qubit ansatz can be compared on the same task by editing a single integer.

\begin{figure}[H]
\centering
\includegraphics[width=\textwidth]{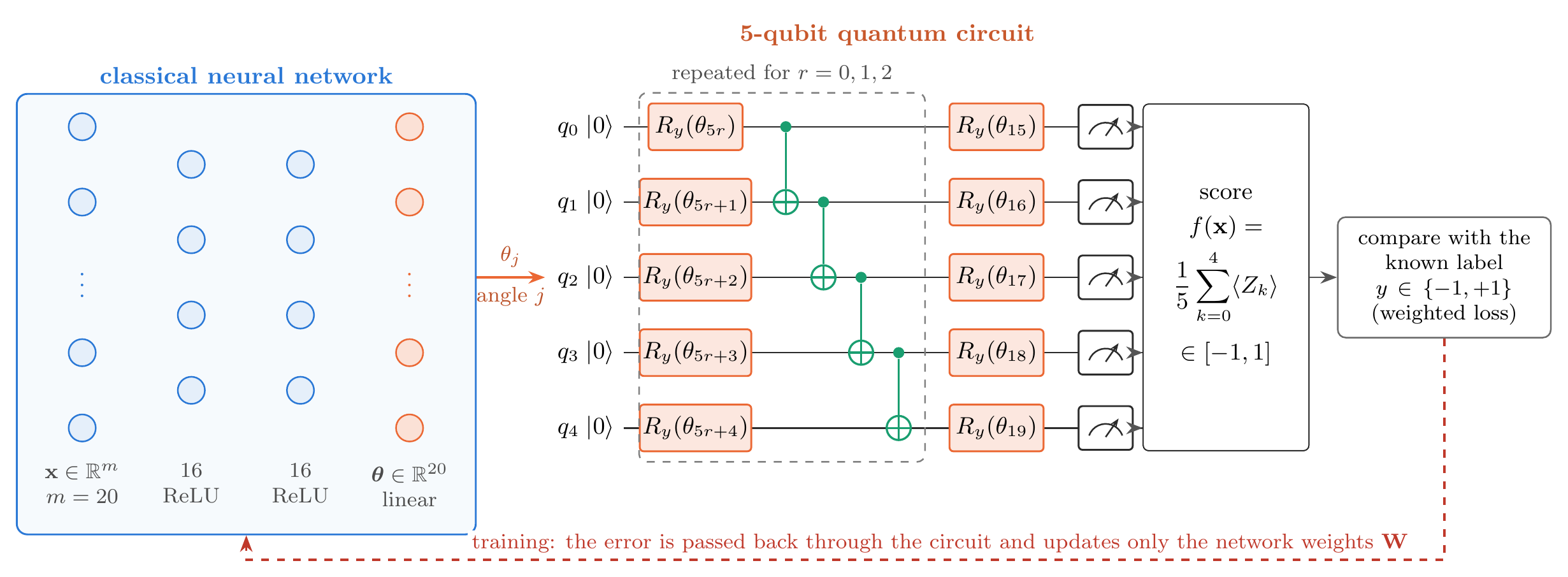}
\caption{\label{fig:hqnn-circuit} The HQNN configuration evaluated in this work. A classical network ($20\to16\to16\to20$, ReLU hidden layers, linear output) maps the $20$ input features to the $20$ $R_y$ angles of a five-qubit RealAmplitudes circuit with linear CNOT entanglement and three repetitions. The score is the average of the five single-qubit $Z$ expectations. During training, the error on the class-weighted loss is passed back through the circuit and updates only the network weights.}
\end{figure}

\textit{Configuration evaluated in this work.} The HQNN reported in Sec.~\ref{sec:hqnn-results} is the simplest instance of this design (Fig.~\ref{fig:hqnn-circuit}). The FFNN maps the $d_{\mathrm{in}}=20$ features through two hidden layers of $16$ ReLU units to a linear output of $D=20$ angles ($948$ trainable weights and biases in total). The circuit is a RealAmplitudes ansatz on $N=5$ qubits with linear CNOT entanglement and three repetitions: three blocks of $R_y$ rotations each followed by a CNOT chain, and a final $R_y$ layer, giving $(3+1)\times5=20$ rotation angles and no other gates. The data enter the circuit only through these angles. The readout is the register-averaged local observable $\hat{O}=\frac{1}{5}\sum_{k=1}^{5}Z_k$, so that $\hat{y}\in[-1,1]$ plays the same role as the QNN score of Sec.~\ref{sec:postprocessing}.
This differs from the re-uploading classifier of Ref.~\cite{perezsalinas2020reuploading}, where a fixed encoding gate alternates with independently trainable blocks and the output is a Fourier series in $\mathbf{x}$ with spectrum fixed by the encoding generators \cite{schuld2021expressive}. Here there is no separate encoding and no free circuit parameter: every angle is a learned function of the input, so the encoding is itself learned and the spectral restriction does not bound the model in $\mathbf{x}$. The cost is that expressivity is relocated to the classical network, with the quantum layer supplying a structured nonlinearity on $\mathbb{R}^{D}$. Whether that nonlinearity does useful work beyond what a classical network of comparable size would do on its own is an empirical question that the comparison in Sec.~\ref{sec:hqnn-results} bounds but does not settle (Sec.~\ref{sec:discussion}).

Since the circuit exposes no free parameters, the trainable set is $(W,b)$ alone: the parameter-shift rule \cite{mitarai2018qcl,schuld2019gradients} supplies $\partial\hat{y}/\partial\theta_k$ exactly, and the chain rule contracts it with the network Jacobian, $\partial\mathcal{L}/\partial W = (\partial\mathcal{L}/\partial\hat{y})\sum_k (\partial\hat{y}/\partial\theta_k)(\partial\theta_k/\partial W)$. Local observables retain trainable gradients at shallow depth, whereas global ones induce cost-function-dependent barren plateaus \cite{cerezo2021costfunction}. Because each sample induces different angles, the shifted circuits cannot be amortized across a mini-batch as they are in the standard variational setting, so a batch of size $B$ costs $2DB$ circuit evaluations per step; runtime is governed by $D$, not $N$. The network is trained with Adam~\cite{kingma2015adam} for $300$ epochs on an ideal (noiseless) simulator, under each of the four class-weighted losses of Sec.~\ref{sec:loss}; the wieghted mean squared error (WMSE) and the weighted binary cross entropy (WBCE) turned out to be the best models, for ANOVA and PCA respectively, and are the ones reported here.

Finally, the decoupling is of the data dimension, not of model capacity. The terminal layer holds $O(D)$ weights, so enlarging the circuit enlarges the classical model, and the HQNN's total trainable parameter count ($948$) is of the same order as that of the classical NN1 baseline of Sec.~\ref{sec:methods:classical}.
\subsubsection{Variational Ansatz and Loss Functions}
\label{sec:loss}
The state prepared by the feature map is processed by a variational circuit $W(\bm{\theta})$, a parametrized sequence of single-qubit rotations interleaved with entangling layers whose angles $\bm{\theta}$ constitute the trainable parameters of the model. (In the hybrid extension of Sec.~\ref{sec:hqnn}, the FFNN-generated circuit $\prod_l U_l(\boldsymbol{\theta}_l)$ takes the place of the feature map and $W(\bm{\theta})$ together.) Several standard families are available for this role, differing in the rotation axes employed and in the connectivity of the entangling blocks: the hardware-efficient \textit{EfficientSU2} ansatz, built from general $SU(2)$ rotations; the \textit{TwoLocal} family, in which arbitrary rotation and entanglement gates may be specified; and the \textit{RealAmplitudes} ansatz, comprising $R_y$ rotations and CNOT entanglers, which prepares states with purely real amplitudes. We adopt the last of these with pairwise entanglement and three repetitions. The restriction to real amplitudes entails no loss of representational power for binary classification, in which the decision function depends only on a real expectation value, while affording a shallower circuit than a general $SU(2)$ construction of equal depth.
Additional repetitions enlarge the expressible function class and would, in principle, admit a better model, but each one appends a further layer of $R_y$ rotations and a complete round of pairwise CNOTs. The cost is incurred twice over. The optimization lengthens, since the derivative-free method employed here (Sec.~\ref{sec:qnn-results}) must explore a search space that grows with every added layer; and the circuit accumulates two-qubit gate error, which is the dominant noise channel on current superconducting devices and is compounded by the routing overhead that pairwise connectivity incurs on a heavy-hex coupling map. Deeper randomly initialized circuits are additionally liable to a flattening of the loss landscape. We use three repetitions and did not pursue the depth systematically.
The parameters $\bm{\theta}$ are updated entirely classically: the quantum circuit is evaluated only to furnish the expectation values entering the objective, which is then minimized on a classical optimizer. The choice of that objective is not incidental here, since the financial data considered in this work is strongly imbalanced, with suspicious samples constituting only about $20\%$ of the total. To account for this, we employ four loss functions, each furnishing a distinct optimization route to $\bm{\theta}$. Three share the same weighted structure and are summarized compactly in Table~\ref{tab:loss-functions}; the fourth, MHINGE, recurs later in the Ising-loss discussion of Appendix~\ref{sec:ising-loss} and is given its own equation, Eq.~\eqref{eq:mhinge}, below.
\begin{table}[H]
\centering
\caption{The three weighted loss functions sharing the class-weighted structure $w_i = N/2N_{c(i)}$, where $N$ is the total number of samples and $N_{c(i)}$ the number belonging to the class of sample $i$. Here $t_i \in \{-1,+1\}$ is the true label and $p_i \in [-1,1]$ the measured expectation value for sample $i$.}
\label{tab:loss-functions}
\begin{tabular}{p{0.13\linewidth}p{0.75\linewidth}}
\toprule
\textbf{Loss} & \textbf{Definition} \\
\midrule
WMSE & $\mathcal{L}_{\text{wmse}} = \dfrac{\sum_i w_i (t_i-p_i)^2}{\sum_i w_i}$ \\[1.5ex]
WBCE & $\mathcal{L}_{\text{wbce}} = -\dfrac{\sum_i w_i \left[y_i \log \hat{y}_i + (1-y_i)\log(1-\hat{y}_i)\right]}{\sum_i w_i}$, with $\hat{y}_i = \frac{1+p_i}{2} \in [0,1]$ and $y_i = \frac{1+t_i}{2} \in \{0,1\}$ \\[1.5ex]
WHINGE & $\mathcal{L}_{\text{whinge}} = \dfrac{\sum_i w_i (1 - t_i p_i)}{\sum_i w_i}$ \\
\bottomrule
\end{tabular}
\end{table}
The fourth objective, \textit{Modified Hinge} (MHINGE), is
\begin{equation}
\mathcal{L}_{\text{mhinge}} = \sum_i \left[\left(1 - t_i p_i\right) + \left(1 - p_i\right)\left(1 + t_i\right)\right]
\label{eq:mhinge}
\end{equation}
Both hinge objectives are written without the rectifier $\max(0,\cdot)$ of the canonical form. This entails no loss of generality here: the observable expectation value is confined to $p_i \in [-1,1]$, so that $t_i p_i \leq 1$ and the argument $1 - t_i p_i$ is non-negative for every sample. The rectifier is therefore inactive throughout the accessible range and the expressions above coincide with the clipped hinge loss on that domain. The same bound guarantees the positivity of Eq.~\eqref{eq:mhinge}. A consequence of the restricted range is that the loss attains zero only at $p_i = t_i$, so that the flat margin region characteristic of the unbounded hinge loss is never entered; the objective retains a non-vanishing gradient for all correctly classified samples short of the extremal prediction. This linear structure is not incidental, but follows from the Ising construction detailed in Appendix~\ref{sec:ising-loss}.
Training against four distinct objectives allows the optimal parameters $\bm{\theta}$ to be selected from four independent optimization routes. The objective that furnishes the best model is reported alongside each result (Secs.~\ref{sec:qnn-results} and~\ref{sec:hqnn-results}); for the HQNN it is WMSE.
The class weights encode the requirement that misclassifying a suspicious sample as benign, a false negative, be penalized more heavily than the converse. For an $80/20$ imbalance they evaluate to $w_+ = 2.5$ and $w_- = 0.625$, so that the positive class carries four times the weight of the negative. The effect is to displace the decision boundary toward the minority class, suppressing false negatives and enhancing recall, generally at some cost in precision.
The MHINGE objective achieves a comparable asymmetry by a different mechanism, and without reference to the empirical class frequencies. Its second term vanishes identically for $t_i = -1$ and contributes $2(1-p_i)$ for $t_i = +1$, so that Eq.~\eqref{eq:mhinge} reduces to $3(1-p_i)$ on suspicious samples and to $(1+p_i)$ on benign ones. The positive class is thereby penalized three times as heavily, through a ratio fixed by the functional form rather than inferred from the data. At an $80/20$ imbalance MHINGE consequently applies a weaker correction than WHINGE, and unlike the latter it would not adapt were the imbalance to change. MHINGE is nonetheless retained as one of the four candidate objectives, since the absence of explicit weights makes it a useful check on whether the empirical class weights are doing the work.
\subsubsection{Measurement and Readout}
The circuit output is the expectation value of a Hermitian observable $M$,
\begin{equation}
f_{\boldsymbol{\theta}}(\mathbf{x}) = \langle\psi(\mathbf{x},\boldsymbol{\theta})|\,\mathcal{O}\,|\psi(\mathbf{x},\boldsymbol{\theta})\rangle \in [-1,1].
\end{equation}
We take $\mathcal{O}$ to be a \emph{local} observable, the sum of local $Z$ operators $\frac{1}{n}\sum_q Z_q$. The choice is not incidental: global observables induce barren plateaus even at constant depth, whereas local observables admit trainable gradients for shallow circuits \cite{cerezo2021costfunction}. Under sampling with $N_{\mathrm{shots}}$, $f_{\boldsymbol{\theta}}$ is estimated with standard error $O(N_{\mathrm{shots}}^{-1/2})$, which sets the resolution floor for both the loss and its gradient.
\subsubsection{Classical Post-Processing: Decision Boundary Selection}
\label{sec:postprocessing}
The network output defined above is an expectation value $\hat{y}_i = \langle \mathcal{O} \rangle \in [-1, 1]$ rather than a class label, so a decision threshold $\tau$ must be fixed before a classification can be issued, the predicted label being $\mathrm{sgn}(\hat{y}_i - \tau)$. The natural choice $\tau = 0$ places the boundary at the midpoint of the output range and thereby presumes balanced classes. This is inappropriate for the present data, in which the benign class constitutes $80\%$ of both splits: the weighted objectives of Sec.~\ref{sec:loss} displace the learned decision function toward the minority class during training, but they do not by themselves determine where along the resulting continuum of outputs the operational boundary should lie. We therefore treat the selection of $\tau$ as a distinct classical post-processing stage, applied after training and independently of it, on the training split.
The procedure traces the precision--recall curve by sweeping $\tau$ over the range of observed outputs on the training data, evaluating precision and recall for the suspicious class at each candidate value, and selecting $\tau^{\star}$ according to one of three operational criteria:
\begin{enumerate}[label=(\roman*)]
    \item maximization of the $F_\beta$ score,
        \begin{equation}
            F_\beta = \left(1 + \beta^2\right) \frac{\mathrm{Precision} \cdot \mathrm{Recall}}{\beta^2\,\mathrm{Precision} + \mathrm{Recall}},
            \label{eq:fbeta}
        \end{equation}
        in which $\beta > 1$ weights recall above precision, the limit $\beta = 1$ recovering the unweighted F1 score;
    \item maximization of recall subject to a floor on precision, $\mathrm{Precision} \geq p_{\min}$;
    \item maximization of precision subject to a floor on recall, $\mathrm{Recall} \geq r_{\min}$;
    \item maximization of recall subject to a ceiling on the false-positive rate, $\mathrm{FPR} \leq f_{\max}$, the fraction of benign accounts flagged.
\end{enumerate}
The criteria encode different operational postures. Criterion (i) fixes a smooth trade-off through the single parameter $\beta$ but does not guarantee any particular level of either quantity. Criteria (ii) and (iii) instead impose a hard constraint and optimize the complementary metric within it, which is the more useful formulation when one of the two carries an external requirement. For trade surveillance, the binding constraint is ordinarily on precision: each positive prediction generates an alert that consumes analyst time, so $p_{\min}$ encodes the tolerable false-alert burden under a fixed review capacity, and recall is then to be made as large as that capacity permits. Criterion (iv) expresses the same budget directly in terms of benign accounts, which unlike precision does not depend on the prevalence of suspicious activity. For the QNN and HQNN we report two operating points (Secs.~\ref{sec:qnn-results} and~\ref{sec:hqnn-results}): an \emph{FPR-minimization} threshold under criterion (iv) at a tight budget, $f_{\max}=0.15$, and a \emph{recall-maximization} threshold under the same criterion at a looser budget, $f_{\max}\approx0.25$, which buys additional recall at the cost of more alerts.

For the QNN and HQNN, $\tau^\star$ is selected on the training split alone and applied unchanged to the held-out test set, so no test-set information enters the classifier at any stage. Because $\tau^\star$ is fit on the training distribution, a floor or ceiling is a target rather than a guarantee once the same threshold is carried over to test data.
\subsection{Quantum Support Vector Machine (QSVM)}
\label{sec:methods:qsvm}
Support Vector Machines (SVMs) construct optimal decision hyperplanes in high-dimensional feature spaces by maximizing the classification margin between distinct data distributions \cite{boser1992, scholkopf2002}. In financial trade surveillance, identifying sophisticated market manipulation, such as the pump-and-dump typologies described in Section~\ref{sec:dataset}, requires capturing complex, non-linear feature interactions \cite{whitrow2009, akoglu2015}. While classical kernel tricks, such as the Radial Basis Function (RBF) kernel, implicitly project features into an infinite-dimensional Hilbert space, they suffer from quadratic Gram matrix construction scaling $O(M^2 d)$, where $d$ is the dimensionality of the input feature vector and $M$ is the number of transactional samples, and risk over- or under-fitting isotropic boundaries on the anisotropic financial data distributions produced by our synthetic generator \cite{rahimi2007, kubler2021inductivebias}.
Quantum Support Vector Machines (QSVMs) address this by leveraging quantum feature maps to embed classical input vectors $\mathbf{x} \in \mathbb{R}^d$ into state vectors within a $2^n$-dimensional complex quantum Hilbert space $\mathcal{H} = (\mathbb{C}^2)^{\otimes n}$, where $n$ is the number of qubits \cite{rebentrost2014, havlicek2019supervised, schuld2019featurehilbert}. In general $n$ need not equal $d$, but the construction adopted here encodes one feature per qubit, so $n = d$ throughout this work (Sec.~\ref{sec:methods:qsvm:featuremap}). By exploiting quantum entanglement and superposition, quantum kernels evaluate non-linear similarities that are classically intractable to reproduce with a fixed-form kernel, offering an alternative inductive bias for financial crime detection \cite{havlicek2019supervised, kubler2021inductivebias}. The remainder of this subsection develops the general quantum-kernel formalism (Sec.~\ref{sec:methods:qsvm:formulation}), the specific feature map and entanglement structure adopted for the pump-and-dump task (Sec.~\ref{sec:methods:qsvm:featuremap}), and the two operating regimes, standard and threshold-changing, under which the resulting classifier is evaluated (Sec.~\ref{sec:methods:qsvm:comparison}).
\subsubsection{Quantum Kernel Formulation}
\label{sec:methods:qsvm:formulation}
\begin{figure}[H]
    \centering
    \begin{quantikz}
        \lstick{$\ket{0}$} & \gate[wires=4, style={fill=teal!25}]{U_{\Phi}(x_i)} & \gate[wires=4, style={fill=violet!20}]{U_{\Phi}^\dagger(x_j)} & \meter{} \\
        \lstick{$\ket{0}$} & & & \meter{} \\
        \lstick{$\ket{0}$} & & & \meter{} \\
        \lstick{$\ket{0}$} & & & \meter{}
    \end{quantikz}
    \caption{Compute-uncompute (Loschmidt-echo) circuit used to evaluate the fidelity quantum kernel of Eq.~\eqref{eq:qsvm_fqk} between two data points $\mathbf{x}_i$ and $\mathbf{x}_j$. The feature-map unitary $U_{\Phi}(\mathbf{x}_i)$ prepares $|\Phi(\mathbf{x}_i)\rangle$ from the reference state $\ket{0}^{\otimes n}$; its inverse $U_{\Phi}^{\dagger}(\mathbf{x}_j)$ is then applied for the second data point, and all $n$ qubits are measured in the computational basis. The frequency of the all-zero outcome over $N_{\text{shots}}$ repetitions is an unbiased estimator of $K_{\text{FQK}}(\mathbf{x}_i, \mathbf{x}_j) = |\langle 0^n | U_{\Phi}^{\dagger}(\mathbf{x}_j) U_{\Phi}(\mathbf{x}_i) | 0^n \rangle|^2$, requiring no ancilla qubit or destructive SWAP test.}
    \label{fig:fqk_circuit}
\end{figure}
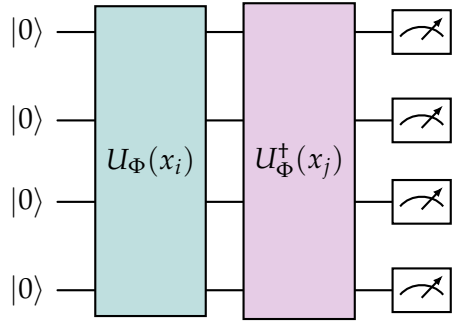

Let $\mathcal{D} = \{(\mathbf{x}_i, y_i)\}_{i=1}^{M}$ denote the training set of $M$ accounts, where $\mathbf{x}_i \in \mathbb{R}^d$ is the $20$-dimensional ANOVA or PCA representation of the account (Sec.~\ref{sec:preprocessing}) and $y_i \in \{-1, +1\}$ is the target label ($-1$ for Benign, $+1$ for Suspicious). The dual convex optimization formulation of the soft-margin Support Vector Classifier maximizes
\begin{equation}
\label{eq:qsvm_svm_dual}
\max_{\boldsymbol{\alpha}} \sum_{i=1}^{M} \alpha_i - \frac{1}{2} \sum_{i=1}^{M} \sum_{j=1}^{M} \alpha_i \alpha_j y_i y_j K(\mathbf{x}_i, \mathbf{x}_j)
\end{equation}
subject to $0 \le \alpha_i \le C$ for all $i \in \{1, \dots, M\}$ and $\sum_{i=1}^{M} \alpha_i y_i = 0$, where $\boldsymbol{\alpha} = (\alpha_1, \dots, \alpha_M)^T$ are Lagrange multipliers, $C > 0$ is the soft-margin regularization hyperparameter, and $K(\mathbf{x}_i, \mathbf{x}_j)$ evaluates an inner product in feature space. A QSVM is precisely an instance of Eq.~\eqref{eq:qsvm_svm_dual} in which $K$ is estimated from a quantum circuit rather than computed in closed form. We characterize four such kernel families below; the practical estimation procedure and the algorithmic contributions of this work are described immediately after.
\paragraph{1. Standard Fidelity Quantum Kernel (FQK).}
Given a parameterized quantum circuit (state encoding map) $U_{\Phi}(\mathbf{x})$ acting on $n$ qubits, a classical input $\mathbf{x} \in \mathbb{R}^d$ is mapped to a pure quantum state
\begin{equation}
\label{eq:qsvm_state_encoding}
|\Phi(\mathbf{x})\rangle = U_{\Phi}(\mathbf{x}) |0\rangle^{\otimes n}.
\end{equation}
The Fidelity Quantum Kernel evaluates the squared transition amplitude between the states encoding two data points $\mathbf{x}_i$ and $\mathbf{x}_j$ \cite{havlicek2019supervised, schuld2019featurehilbert}:
\begin{equation}
\label{eq:qsvm_fqk}
K_{\text{FQK}}(\mathbf{x}_i, \mathbf{x}_j) = \left| \langle \Phi(\mathbf{x}_i) \mid \Phi(\mathbf{x}_j) \rangle \right|^2 = \mathrm{Tr}\left( \rho(\mathbf{x}_i) \rho(\mathbf{x}_j) \right),
\end{equation}
where $\rho(\mathbf{x}) = |\Phi(\mathbf{x})\rangle \langle \Phi(\mathbf{x})|$. On hardware or a shot-based simulator, $K_{\text{FQK}}(\mathbf{x}_i, \mathbf{x}_j)$ is obtained by applying the overlap unitary $U_{\Phi}^{\dagger}(\mathbf{x}_j) U_{\Phi}(\mathbf{x}_i)$ to $|0\rangle^{\otimes n}$ and estimating the probability of returning to the all-zero bitstring over $N_{\text{shots}}$ circuit executions. It serves here as the reference formulation: the kernel actually instantiated in our QSVM experiments is its one-body projected counterpart, introduced below, and the remaining families are characterized for theoretical completeness and to motivate future directions.
Figure~\ref{fig:fqk_circuit} illustrates the compute-uncompute, or Loschmidt-echo, protocol used to evaluate $K_{\text{FQK}}$ in practice. This direct-estimation circuit was introduced concurrently by Havl{\'i}{\v{c}}ek et al.~\cite{havlicek2019supervised} and by Schuld and Killoran~\cite{schuld2019featurehilbert} as an alternative to the ancilla-based SWAP test~\cite{buhrman2001swaptest} for estimating a fidelity-type overlap: rather than routing both states through a shared reference qubit, the feature-map unitary for one data point is immediately followed by the \emph{inverse} feature-map unitary for the other, so that the desired transition amplitude is read off directly from the probability of returning to the reference state.
\paragraph{2. Projected Quantum Kernel (PQK).}
As $n$ grows, fidelity kernels concentrate toward zero for almost all distinct pairs, $K_{\text{FQK}}(\mathbf{x}_i, \mathbf{x}_j) \to 0$ for $i \neq j$, an exponential concentration of measure that drives the Gram matrix toward the identity and degrades generalization \cite{huang2021powerofdata}. Projected Quantum Kernels avoid this by projecting the $n$-qubit state $\rho(\mathbf{x})$ onto single-qubit reduced density matrices,
\begin{equation}
\label{eq:qsvm_1rdm}
\langle P^{(k)} \rangle_{\mathbf{x}} = \mathrm{Tr}\left( P \cdot \mathrm{Tr}_{\overline{k}}\big(\rho(\mathbf{x})\big) \right), \qquad P \in \{X, Y, Z\},
\end{equation}
where $\mathrm{Tr}_{\overline{k}}(\cdot)$ traces out every qubit except $k$. The resulting expectation values are then compared classically through a radial basis kernel \cite{huang2021powerofdata}:
\begin{equation}
\label{eq:qsvm_pqk}
\begin{split}
K_{\text{PQK}}(\mathbf{x}_i, \mathbf{x}_j) = \exp\Big( -\gamma \sum_{k=1}^{n} \sum_{P \in \{X, Y, Z\}} \\
\big( \langle P^{(k)} \rangle_{\mathbf{x}_i} - \langle P^{(k)} \rangle_{\mathbf{x}_j} \big)^2 \Big),
\end{split}
\end{equation}
with $\gamma > 0$ a bandwidth hyperparameter. Projecting to local observables retains non-linear entanglement effects while avoiding Hilbert-space saturation \cite{huang2021powerofdata}.

\begin{figure}[H]
    \centering
    \begin{tikzpicture}[
        >=Stealth,
        wire/.style={line width=0.6pt},
        ubox/.style={draw, rounded corners=2pt, minimum width=2.3cm, minimum height=1.6cm, align=center},
        rbox/.style={draw, rounded corners=4pt, fill=white, minimum height=0.75cm, inner sep=4pt}
    ]
    % Register for x_i
    \node[ubox, fill=teal!25] (Ui) at (2.2,2.5) {$U_{\Phi}(\mathbf{x}_i)$};
    \node at (0.25,3.1) {$\ket{0}$};
    \node at (0.25,2.55) {$\vdots$};
    \node at (0.25,1.9) {$\ket{0}$};
    \draw[wire] (0.6,3.1) -- (Ui.west |- 0,3.1);
    \draw[wire] (0.6,1.9) -- (Ui.west |- 0,1.9);
    \node[rbox] (ri) at (7.6,3.1) {$\rho_{(1)}(\mathbf{x}_i)$};
    \draw[wire] (Ui.east |- 0,3.1) -- (ri.west);
    % Register for x_j
    \node[ubox, fill=teal!40] (Uj) at (2.2,0.2) {$U_{\Phi}(\mathbf{x}_j)$};
    \node at (0.25,0.8) {$\ket{0}$};
    \node at (0.25,0.25) {$\vdots$};
    \node at (0.25,-0.4) {$\ket{0}$};
    \draw[wire] (0.6,0.8) -- (Uj.west |- 0,0.8);
    \draw[wire] (0.6,-0.4) -- (Uj.west |- 0,-0.4);
    \node[rbox] (rj) at (5.4,0.8) {$\rho_{(1)}(\mathbf{x}_j)$};
    \draw[wire] (Uj.east |- 0,0.8) -- (rj.west);
    % Kernel assembly
    \node[rbox, fill=gray!20, minimum width=5.6cm, minimum height=0.9cm] (K) at (6.1,-1.5)
        {$K_{\text{1BPQK}}(\mathbf{x}_i,\mathbf{x}_j) = \mathrm{Tr}\big[\rho_{(1)}(\mathbf{x}_i)\,\rho_{(1)}(\mathbf{x}_j)\big]$};
    \draw[->, wire] (rj.south) -- (rj.south |- K.north);
    \draw[->, wire] (ri.south) -- (ri.south |- K.north);
    \end{tikzpicture}
    \caption{One-body projected quantum kernel used for every QSVM result in this work, after Ref.~\cite{rodriguezgrasa2025neural}. Each data point is encoded by the Pauli-$ZZ$ feature map $U_{\Phi}$ of Sec.~\ref{sec:methods:qsvm:featuremap} on an $n$-qubit register initialised in $\ket{0}^{\otimes n}$. Only the top qubit is measured, in the $X$, $Y$ and $Z$ bases, giving its reduced density matrix $\rho_{(1)}$, the remaining $n-1$ qubits being traced out. The kernel entry is then assembled classically from the two one-qubit states, $K_{\text{1BPQK}}(\mathbf{x}_i,\mathbf{x}_j) = (1 + \vec{r}_i \cdot \vec{r}_j)/2$, so no overlap circuit between data points is executed.}
    \label{fig:qsvm_onebody_circuit}
\end{figure}
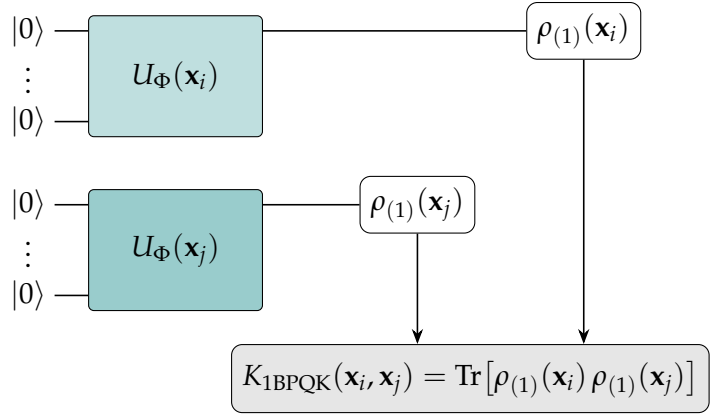

\textbf{One-body (density-matrix) projected quantum kernel.} The kernel used for every QSVM result in this paper is the linear, one-body form of this projection advocated by Rodriguez-Grasa, Ban and Sanz~\cite{rodriguezgrasa2025neural}, shown schematically in Fig.~\ref{fig:qsvm_onebody_circuit}. Each data point $\mathbf{x}_i$ is encoded by the feature map $U_{\Phi}(\mathbf{x}_i)$ of Eq.~\eqref{eq:qsvm_state_encoding}, and only the first (top) qubit of the register is retained: its one-qubit reduced density matrix $\rho_{(1)}(\mathbf{x}_i) = \mathrm{Tr}_{\overline{1}}\big(\rho(\mathbf{x}_i)\big)$ is obtained by tracing out the remaining $n-1$ qubits, as in Ref.~\cite{rodriguezgrasa2025neural}. Measuring this qubit alone in the $X$, $Y$ and $Z$ bases gives its Bloch vector $\vec{r}_i = \big(\langle X\rangle, \langle Y\rangle, \langle Z\rangle\big)_i$, from which the training and test kernel matrices are assembled classically as
\begin{equation}
\label{eq:qsvm_onebody}
K_{\text{1BPQK}}(\mathbf{x}_i, \mathbf{x}_j) = \mathrm{Tr}\!\left[\rho_{(1)}(\mathbf{x}_i)\,\rho_{(1)}(\mathbf{x}_j)\right] = \frac{1 + \vec{r}_i \cdot \vec{r}_j}{2}.
\end{equation}
Because each data point is measured once, in three Pauli bases, rather than once for every pair of points, building the full kernel matrix requires $O(M)$ circuit executions in place of the $O(M^2)$ overlap circuits of Fig.~\ref{fig:fqk_circuit}; and because the kernel is built from local rather than global quantities, it does not suffer the exponential concentration that drives $K_{\text{FQK}}$ toward the identity as $n$ grows \cite{huang2021powerofdata,rodriguezgrasa2025neural}.
\paragraph{3. Quantum Kernel Alignment (QKA).}
Rather than fixing the embedding unitary, Quantum Kernel Alignment introduces variational parameters $\boldsymbol{\theta}$ into $U_{\Phi}(\mathbf{x}; \boldsymbol{\theta})$ \cite{hubregtsen2022}, optimized prior to SVM fitting by maximizing the Kernel-Target Alignment against the ideal label matrix $\mathbf{Y} = \mathbf{y}\mathbf{y}^T$ \cite{hubregtsen2022}:
\begin{equation}
\label{eq:qsvm_kta}
\begin{split}
S_{TA}(K_{\boldsymbol{\theta}}, \mathbf{Y}) &= \frac{\langle K_{\boldsymbol{\theta}}, \mathbf{Y} \rangle_F}{\|K_{\boldsymbol{\theta}}\|_F \|\mathbf{Y}\|_F} \\
&= \frac{\sum_{i,j=1}^{M} K_{\boldsymbol{\theta}}(\mathbf{x}_i, \mathbf{x}_j) y_i y_j}{\sqrt{\left(\sum_{i,j=1}^{M} K_{\boldsymbol{\theta}}^2(\mathbf{x}_i, \mathbf{x}_j)\right)\left(\sum_{i,j=1}^{M} y_i^2 y_j^2\right)}},
\end{split}
\end{equation}
where $\langle \cdot, \cdot \rangle_F$ and $\|\cdot\|_F$ are the Frobenius inner product and norm. Maximizing $S_{TA}$ pulls same-class states closer together in kernel space while pushing opposite-class states apart \cite{hubregtsen2022}.
\paragraph{4. Neural Quantum Embedding (NQE).}
To scale quantum kernels to tabular data with many engineered features without an impractically large qubit register, a hybrid Neural Quantum Embedding composes a classical trainable backbone $f_{\boldsymbol{\theta}}: \mathbb{R}^d \to \mathbb{R}^n$ with the quantum encoding circuit \cite{lloyd2020, hur2024}:
\begin{equation}
\label{eq:qsvm_nqe}
|\Phi(\mathbf{x}; \boldsymbol{\theta})\rangle = U_{\Phi}\big(f_{\boldsymbol{\theta}}(\mathbf{x})\big) |0\rangle^{\otimes n}.
\end{equation}
The network $f_{\boldsymbol{\theta}}$ is trained end to end against a kernel-target-alignment or contrastive objective, jointly optimizing classical dimensionality reduction and quantum feature mapping \cite{hur2024}.
\paragraph{Practical kernel matrix estimation and algorithmic contributions.}
\label{sec:methods:qsvm:practical}
The kernel matrices are obtained in two stages: a quantum stage that estimates one Bloch vector per account, and a classical stage that assembles the Gram matrices from these vectors. In the quantum stage, the feature-map circuit $U_{\Phi}(\mathbf{x})$ of Sec.~\ref{sec:methods:qsvm:featuremap} is constructed once as a parameterized template and bound to each feature vector by parameter assignment, rather than rebuilt for every account. Each bound circuit is paired with the three single-qubit observables $X$, $Y$ and $Z$ on the top qubit (the identity acting on the remaining $n-1$ qubits), and all circuits of a split are submitted to the estimator in a single batched call at a target precision of $1/\sqrt{N_{\text{shots}}}$. Building the kernels therefore requires $M + M_{\text{test}}$ circuit evaluations, one per account, in place of the $O(M^2)$ train-train and $M_{\text{test}} \times M$ test-train overlap circuits of Fig.~\ref{fig:fqk_circuit}. The estimated Bloch vectors are not renormalized: since the top qubit is generically entangled with the rest of the register, its reduced state is mixed and $\|\vec{r}_i\| < 1$ carries information about that entanglement rather than shot noise, and Eq.~\eqref{eq:qsvm_onebody} holds for mixed states. In the classical stage, the train-train and test-train kernel matrices are evaluated from Eq.~\eqref{eq:qsvm_onebody} as matrix products of the stored Bloch vectors, and the diagonal of the training kernel is set to unity. Because $K_{\text{1BPQK}}$ is an inner product of the vectors $(1,\vec{r}_i)/\sqrt{2}$, its Gram matrix is positive semi-definite (PSD) by construction, as required by the dual formulation of Eq.~\eqref{eq:qsvm_svm_dual}, even when the Bloch vectors are estimated from finitely many shots, and setting the diagonal to unity adds only a non-negative diagonal term that preserves this property. As a numerical safeguard, we nevertheless eigendecompose the training-kernel matrix $\hat{K} = V \Lambda V^T$, clip any negative eigenvalues to zero, $\Lambda_{\text{PSD}} = \max(\Lambda, 0)$, and reconstruct $K_{\text{PSD}} = V \Lambda_{\text{PSD}} V^T$, the nearest PSD matrix to $\hat{K}$ in Frobenius norm, which leaves an already PSD matrix unchanged.
Table~\ref{tab:qsvm_kernel_comparison} places the one-body projected kernel used in this work alongside the classical and quantum alternatives characterized above; this comparison is theoretical, and no classical RBF baseline was trained as part of this study.
\begin{table}[H]
\centering
\caption{Comparison of classical and quantum kernel formulations.}
\label{tab:qsvm_kernel_comparison}
\small
\resizebox{\textwidth}{!}{%
\begin{tabular}{p{2.3cm}|p{4.2cm}|p{3.2cm}|p{4.5cm}}
\hline
\textbf{Kernel Type} & \textbf{Mathematical Formulation} & \textbf{Evaluation Complexity} & \textbf{Inductive Bias \& Characteristics} \\
\hline
\textbf{Linear SVM} & $K(\mathbf{x}_i, \mathbf{x}_j) = \mathbf{x}_i^T \mathbf{x}_j$ & $O(d)$ classical & Fits linear hyperplanes; fails on non-linear pump-and-dump signals \cite{boser1992}. \\
\hline
\textbf{RBF (Gaussian) SVM} & $K(\mathbf{x}_i, \mathbf{x}_j) = \exp\left(-\gamma \|\mathbf{x}_i - \mathbf{x}_j\|^2\right)$ & $O(d)$ classical & Infinite-dimensional isotropic similarity; sensitive to local Euclidean distance \cite{scholkopf2002}. \\
\hline
\textbf{Fidelity QSVM (FQK)} & $K(\mathbf{x}_i, \mathbf{x}_j) = \left|\langle \Phi(\mathbf{x}_i) \mid \Phi(\mathbf{x}_j) \rangle\right|^2$ & $O(N_{\text{shots}} \cdot d_m \cdot n^2)$ quantum & Non-linear inner products in $2^n$ Hilbert space; suffers from measure concentration \cite{havlicek2019supervised, huang2021powerofdata}. \\
\hline
\textbf{Projected QSVM (PQK)} & $K(\mathbf{x}_i, \mathbf{x}_j) = \exp\left(-\gamma \sum \Delta \langle P \rangle^2\right)$;\newline one-body: $K_{\text{1BPQK}}(\mathbf{x}_i, \mathbf{x}_j) = \mathrm{Tr}\left[\rho_{(1)}(\mathbf{x}_i)\,\rho_{(1)}(\mathbf{x}_j)\right] = \frac{1 + \vec{r}_i \cdot \vec{r}_j}{2}$ &$O(N_{\text{shots}} \cdot n)$ quantum & 1-body projection spaces; mitigates Hilbert space saturation while retaining entanglement \cite{huang2021powerofdata,rodriguezgrasa2025neural}. One-body form (Eq.~\eqref{eq:qsvm_onebody}) used in this work. \\
\hline
\textbf{Aligned QSVM (QKA / NQE)} & $K(\mathbf{x}_i, \mathbf{x}_j; \boldsymbol{\theta}) = K\Big(U_{\Phi}\big(f_{\boldsymbol{\theta}}(\mathbf{x}_i)\big), U_{\Phi}\big(f_{\boldsymbol{\theta}}(\mathbf{x}_j)\big)\Big)$ & Hybrid $O(N_{\text{shots}} \cdot n) +$ classical backpropagation & Data-driven optimization of state embeddings; aligns kernel metrics directly with labels \cite{hubregtsen2022, hur2024}. \\
\hline
\end{tabular}}
\end{table}
Quantum feature maps warp input distances non-linearly through multi-qubit unitary transformations \cite{schuld2019featurehilbert, kubler2021inductivebias}, in contrast to the isotropic Euclidean similarity of a classical RBF kernel (Table~\ref{tab:qsvm_kernel_comparison}). While classical kernel machines require $O(M^2 d)$ scalar operations to build the Gram matrix, fidelity QSVMs require $O(M^2)$ pairwise quantum circuit executions, each sampled over $N_{\text{shots}}$ measurements \cite{havlicek2019supervised}; the one-body projected kernel used here reduces this to $O(M)$ circuit executions, followed by an $O(M^2)$ classical assembly of Eq.~\eqref{eq:qsvm_onebody}. The key theoretical motivation for testing QSVMs on trade surveillance is the structural mismatch between classical isotropic kernels and the anisotropic, multi-scale manipulation patterns produced by our pump-and-dump generator (Sec.~\ref{sec:dataset}): benign and suspicious accounts become non-linearly entangled in raw feature space \cite{fincen2021}, and the non-linear Pauli-$ZZ$ expansion of Sec.~\ref{sec:methods:qsvm:featuremap} maps these cross-feature interactions into a quantum Hilbert space where an optimal linear hyperplane can separate patterns that overwhelm a shallow classical kernel \cite{havlicek2019supervised, hubregtsen2022, hur2024}.
\subsubsection{Feature Map and Entanglement Structure}
\label{sec:methods:qsvm:featuremap}
The expressive power of the QSVM rests on the structure of the feature map $U_{\Phi}(\mathbf{x})$ \cite{havlicek2019supervised}, which is the only circuit in the model and contains no trainable parameters. Classical inputs, the $d = 20$ components of the ANOVA or PCA representation of Sec.~\ref{sec:preprocessing}, each rescaled to $[0,\pi]$, are encoded using the second-order Pauli-$ZZ$ feature map of $d_m$ repeated layers,
\begin{equation}
\label{eq:qsvm_zzfm}
U_{\Phi}(\mathbf{x}) = \prod_{l=1}^{d_m} \left( U_{\phi(\mathbf{x})} H^{\otimes n} \right),
\end{equation}
where $H^{\otimes n}$ prepares an equal superposition over all $n$ qubits and $U_{\phi(\mathbf{x})}$ is the diagonal phase gate
\begin{equation}
\label{eq:qsvm_phasegate}
U_{\phi(\mathbf{x})} = \exp\left( i \sum_{S \subseteq [n]} \phi_S(\mathbf{x}) \prod_{k \in S} Z_k \right),
\end{equation}
with $Z_k$ the Pauli-$Z$ operator on qubit $k$ and $S \subseteq [n] = \{1, \dots, n\}$ indexing interaction subsets. We restrict to single-qubit terms ($|S|=1$) and pairwise entangling terms ($|S|=2$):
\begin{equation}
\label{eq:qsvm_singleenc}
\phi_{\{k\}}(\mathbf{x}) = x_k,
\end{equation}
\begin{equation}
\label{eq:qsvm_pairenc}
\phi_{\{k, l\}}(\mathbf{x}) = (\pi - x_k)(\pi - x_l).
\end{equation}
Equations~\eqref{eq:qsvm_singleenc} and~\eqref{eq:qsvm_pairenc} are the original encoding of Havl{\'i}{\v{c}}ek et al.~\cite{havlicek2019supervised}. This non-linear coupling translates products of classical features, such as wash-trading frequency and trade-size anomaly $z$-scores, into controlled phase shifts, synthesizing multi-particle entanglement across the register \cite{havlicek2019supervised}.
$U_{\phi(\mathbf{x})}$ admits one of three entangling topologies in general: \textbf{linear (pairwise)} nearest-neighbor $ZZ$ gates, $S = \{(k, k+1) \mid 1 \le k \le n-1\}$; \textbf{circular}, which adds a periodic boundary term $S = \{(k, k+1) \mid 1 \le k \le n-1\} \cup \{(n, 1)\}$; and \textbf{full (all-to-all)}, $S = \{(k, l) \mid 1 \le k < l \le n\}$. The results reported in Sec.~\ref{sec:qsvm-results} use only the linear (pairwise) topology at $d_m = 2$, in which the nearest-neighbour $ZZ$ gates are applied in two alternating layers, first on the pairs with odd $k$ and then on those with even $k$, so that gates on disjoint pairs act in parallel; because the $ZZ$ phase gates are diagonal and commute, this arrangement implements the same unitary as the sequential linear ordering at reduced circuit depth. The circular and full topologies are defined here for completeness but were not run in this study. Each account is encoded with one qubit per feature, so the $d = 20$ features of each representation (ANOVA or PCA; Sec.~\ref{sec:preprocessing}) are mapped onto a $20$-qubit register. The one-body expectations entering Eq.~\eqref{eq:qsvm_onebody} were estimated with $N_{\text{shots}} = 4096$ under the noise models of two IBM fake backends, \texttt{FakeFez} and \texttt{FakeGuadalupeV2}, so every QSVM result includes simulated device noise; no physical-device execution is included. For both backends, only the device noise model (\texttt{NoiseModel.from\_backend}) was imported into a matrix-product-state Aer simulator, without the device coupling map, so the $20$-qubit circuits are not routed onto the physical topology of either device.
\subsubsection{Standard versus Threshold-Changing QSVM}
\label{sec:methods:qsvm:comparison}
Once the PSD-corrected one-body projected kernel matrix of Sec.~\ref{sec:methods:qsvm:practical} is available, it is passed as a precomputed Gram matrix to a classical dual SVM solver, which returns a continuous decision score
\begin{equation}
\label{eq:qsvm_decision}
f(\mathbf{x}) = \sum_{i=1}^{M} \alpha_i y_i K(\mathbf{x}_i, \mathbf{x}) - b,
\end{equation}
where $b$ is the bias fixed by the dual solver itself at training time, via the Karush--Kuhn--Tucker (KKT) conditions~\cite{boydvandenberghe2004convex} on the support vectors, and is therefore distinct from the post hoc threshold $\tau$ introduced below. We evaluate this classifier under two distinct operating regimes.
\paragraph{Standard QSVM.} The conventional decision rule assigns the predicted label $\mathrm{sgn}(f(\mathbf{x}))$, placing the operating point at the zero level set of Eq.~\eqref{eq:qsvm_decision}. This is the default behaviour of the dual solver and, as with the untuned QNN of Sec.~\ref{sec:postprocessing}, implicitly presumes that the cost of a false positive and a false negative are comparable and that the classes are roughly balanced, neither of which holds for the $80{:}20$ benign-to-suspicious split used throughout this study (Sec.~\ref{sec:dataset}).
\paragraph{Threshold-Changing QSVM.} Rather than retrain the kernel or the solver, we treat the operating threshold as a separate, post hoc classical stage applied to the same trained score $f(\mathbf{x})$, mirroring the procedure introduced for the QNN in Sec.~\ref{sec:postprocessing}: the precision--recall curve is traced by sweeping a threshold $\tau$ over the range of observed decision scores on the held-out test split, and $\tau^\star$ is fixed according to one of three operational criteria, namely (i) maximization of the $F_\beta$ score (Eq.~\eqref{eq:fbeta}), (ii) maximization of recall subject to a precision floor $\mathrm{Precision} \ge p_{\min}$, or (iii) maximization of precision subject to a recall floor $\mathrm{Recall} \ge r_{\min}$. The predicted label is then $\mathrm{sgn}(f(\mathbf{x}) - \tau^\star)$ in place of $\mathrm{sgn}(f(\mathbf{x}))$. Because $\tau^\star$ is selected without altering $K$, $\boldsymbol{\alpha}$, or $b$, the standard and threshold-changing QSVMs share an identical trained model and differ only in where the decision boundary is drawn along the score axis; the comparison in Sec.~\ref{sec:qsvm-results} therefore isolates the effect of operating-point selection from the effect of kernel or solver choice. We report the $F_\beta$ criterion at $\beta=1.25$ together with the precision-floor criterion at $p_{\min}\in\{0.55,0.60\}$ and the recall-floor criterion at $r_{\min}\in\{0.70,0.75\}$, reflecting the trade surveillance posture in which each positive prediction consumes fixed analyst review capacity. Because $\tau^\star$ is selected on the same $200$ test accounts on which it is evaluated, the threshold-changing results describe the operating points attainable with the trained kernel and should be read as an optimistic upper bound on what a threshold fixed on training data alone would deliver; the standard rule ($\tau=0$) involves no such selection.
% =====================================================================
\section{Results}
\label{sec:results}
\subsection{Evaluation Protocol}
\label{sec:protocol}
Every result in this section is measured on the same $200$ held-out test accounts ($40$ suspicious, $160$ benign; Sec.~\ref{sec:dataset}) after all training, model selection and, unless stated otherwise, threshold selection have been completed on the $800$ training accounts (the QSVM thresholds being the exception). Precision, recall and F1 are reported for the suspicious class, as percentages, together with the false-positive rate (FPR), the percentage of the $160$ benign accounts that are flagged. Because the methods differ in how a continuous score is turned into an alert, each table states its operating point: the classical models use the threshold that maximises training out-of-fold F1 (Sec.~\ref{sec:methods:classical}); the QKNN issues labels directly from its vote and has no threshold; the QSVM is reported at its default threshold and at the post hoc criteria of Sec.~\ref{sec:methods:qsvm:comparison}; and the HQNN at two thresholds fixed on its training scores (Sec.~\ref{sec:postprocessing}). For reference, a rule that flags every account attains $20\%$ precision, $100\%$ recall and $33.3\%$ F1, and a rule that flags none attains $80\%$ accuracy with zero recall.

\subsection{Classical Baselines}
\label{sec:classical-results}
Table~\ref{tab:classical} reports the eight classical configurations.
\begin{table}[H]
\centering
\caption{Classical baselines on the $200$ test accounts. The threshold $\tau^\star$ maximises pooled out-of-fold F1 on the training accounts and is frozen before testing. ROC AUC and average precision (AP) are threshold-free. ``F1 @ 0.5'' evaluates the same test scores at the default threshold. The seed column gives the mean $\pm$ standard deviation of test F1 over ten training seeds with the selected hyperparameters (logistic regression is deterministic and was run once). Metrics are for the suspicious class, in \%.}
\label{tab:classical}
\resizebox{\textwidth}{!}{%
\begin{tabular}{llccccccc|c}
\toprule
\textbf{Features} & \textbf{Model} & $\tau^\star$ & \textbf{Precision} & \textbf{Recall} & \textbf{F1} & \textbf{ROC AUC} & \textbf{AP} & \textbf{F1 @ 0.5} & \textbf{F1, 10 seeds} \\
\midrule
ANOVA & Logistic regression & 0.504 & 36.5 & 47.5 & 41.3 & 0.728 & 0.407 & 41.1 & 41.3 \\
ANOVA & XGBoost             & 0.497 & \textbf{50.8} & 77.5 & \textbf{61.4} & \textbf{0.866} & \textbf{0.505} & 61.4 & $61.4 \pm 0.0$ \\
ANOVA & NN1                 & 0.252 & 47.8 & 80.0 & 59.8 & 0.838 & 0.456 & 35.3 & $58.2 \pm 1.1$ \\
ANOVA & NN3                 & 0.669 & 42.9 & 67.5 & 52.4 & 0.809 & 0.419 & 55.7 & $53.9 \pm 3.5$ \\
\midrule
PCA   & Logistic regression & 0.528 & 37.5 & \textbf{82.5} & 51.6 & 0.812 & 0.443 & 50.8 & 51.6 \\
PCA   & XGBoost             & 0.356 & 35.3 & 60.0 & 44.4 & 0.736 & 0.324 & 34.1 & $44.4 \pm 0.0$ \\
PCA   & NN1                 & 0.272 & 40.0 & 65.0 & 49.5 & 0.798 & 0.431 & 35.3 & $50.3 \pm 2.7$ \\
PCA   & NN3                 & 0.376 & 40.0 & 70.0 & 50.9 & 0.806 & 0.459 & 50.0 & $50.2 \pm 2.5$ \\
\bottomrule
\end{tabular}}
\end{table}
On the ANOVA representation, XGBoost has the highest test F1 of any model in this study, $61.4\%$: it detects $31$ of the $40$ suspicious accounts while flagging $30$ of the $160$ benign ones (FPR $18.8\%$), and it also has the best threshold-free ranking (ROC AUC $0.866$, AP $0.505$). NN1 is close behind ($59.8\%$, catching $32$ of $40$ at FPR $21.9\%$). On the PCA representation the ordering changes: logistic regression is strongest ($51.6\%$), with the highest recall in the table ($82.5\%$) but $55$ false alarms (FPR $34.4\%$), and XGBoost falls to $44.4\%$. No single classical model is best on both representations, and every ANOVA model except logistic regression outperforms its PCA counterpart.

Two further observations bear on the comparison with the quantum models. First, the threshold matters: training-selected thresholds raise test F1 by $24.5$ points for ANOVA NN1 and $14.2$ points for PCA NN1 relative to the default $0.5$, but lower it by $3.2$ points for ANOVA NN3, so a threshold fitted on training data does not always transfer favourably. Second, seed-to-seed variation is up to $3.5$ points of F1 for the neural networks and zero for XGBoost, whose selected configurations use no row or column subsampling. Differences between models of a few points are therefore within the range produced by training randomness alone.

\subsection{QKNN Results}
\label{sec:qknn-results}
The clustering-based QKNN of Sec.~\ref{sec:fidelity_based_clustering} was evaluated on each of the three representations, first on an ideal shot-based simulator and then under the noise model of \texttt{FakeFez}, with the configuration in Table~\ref{tab:qknn_configuration}. A classical $k$-nearest-neighbours classifier on the same split serves as the like-for-like control.
\begin{table}[H]
\centering
\caption{Experimental configuration for the fidelity-based clustering QKNN and the classical KNN control.}
\label{tab:qknn_configuration}\
\resizebox{\textwidth}{!}{%
\begin{tabular}{lll}
\hline
\textbf{Parameter} & \textbf{Quantum KNN} & \textbf{Classical KNN}\\
\hline
Feature-map encoding & ZZ feature map &  Not Relevant \\
Number of features / qubits & 20 & 20\\
Entanglement & Pairwise & Not Relevant \\
Distance method & Compute--uncompute & Minkowski \\
Number of shots & 16{,}248 (ideal); 4096 (\texttt{FakeFez} noise) & Not Relevant \\
Number of neighbours ($k$) & 5 & 5\\
Maximum number of clusters & 5 & Not Relevant \\
Inter-cluster distance threshold & 0.558 & Not Relevant \\
Inter-cluster distance method & Medoid & Not Relevant \\
Voting method & Mean-fidelity-weighted voting & Uniform voting\\
\hline
\end{tabular}}
\end{table}

\begin{table}[H]
\centering
\caption{QKNN on the $200$ test accounts for each representation, on an ideal simulator and under \texttt{FakeFez} noise, against a classical KNN control. TP, FN, FP and TN are counts; the other columns are in \%, for the suspicious class.}
\label{tab:qknn_results}
\resizebox{\textwidth}{!}{%
\begin{tabular}{llccccccccc}
\toprule
\textbf{Model} & \textbf{Features} & \textbf{Acc.} & \textbf{Precision} & \textbf{Recall} & \textbf{F1} & \textbf{TP} & \textbf{FN} & \textbf{FP} & \textbf{TN} \\
\midrule
% QKNN, ideal & UMAP  & 64.5 & 23.7 & 35.0 & 28.3 & 14 & 26 & 45 & 115 \\
QKNN, ideal & PCA   & 65.5 & 24.6 & 35.0 & 28.9 & 14 & 26 & 43 & 117 \\
QKNN, ideal & ANOVA & 74.0 & 38.5 & 50.0 & 43.5 & 20 & 20 & 32 & 128 \\
\midrule
% QKNN, \texttt{FakeFez} & UMAP  & 64.0 & 25.0 & 40.0 & 30.8 & 16 & 24 & 48 & 112 \\
QKNN, \texttt{FakeFez} & PCA   & 65.5 & 28.4 & 47.5 & 35.5 & 19 & 21 & 48 & 112 \\
QKNN, \texttt{FakeFez} & ANOVA & 72.5 & 37.3 & \textbf{55.0} & \textbf{44.4} & 22 & 18 & 37 & 123 \\
\midrule
Classical KNN & ANOVA & 76.5 & 39.4 & 32.5 & 35.6 & 13 & 27 & 20 & 140 \\
Classical KNN & PCA & 76.0 & 36.7 & 27.5 & 31.4 & 11 & 29 & 19 & 141\\
\bottomrule
\end{tabular}}
\end{table}
The representation dominates the result. On ANOVA features the QKNN detects $20$ of the $40$ suspicious accounts on the ideal simulator (F1 $43.5\%$), against $14$ on PCA (F1 $28.9\%$), so the choice of representation moves F1 by about $15$ points, far more than any other factor examined here. On ANOVA the QKNN also exceeds the classical KNN control ($k=5$, Minkowski distance, uniform voting, Table~\ref{tab:qknn_configuration}) by $7.9$ points of F1 on the ideal simulator: it catches $7$ more suspicious accounts ($20$ against $13$) at the cost of $12$ more false alarms ($32$ against $20$), trading a slightly lower precision for a substantially higher recall. On PCA the ideal-simulator QKNN is slightly below the classical control ($28.9$ against $31.4\%$), while the noisy QKNN is above it ($35.5\%$).

Simulated device noise did not degrade the QKNN. Under the \texttt{FakeFez} noise model, recall rises on both representations, by $2$ (ANOVA) and $5$ (PCA) detected accounts, and F1 rises with it, reaching $44.4\%$ on ANOVA. We do not interpret this as noise improving the classifier: the noisy runs also used a different shot budget ($4096$ against $16{,}248$), and the changes amount to a handful of accounts on a $40$-account minority class, of the same order as the $2.5$-point granularity of recall noted in Sec.~\ref{sec:dataset}. What the comparison does establish is that fidelity estimates perturbed by a realistic noise model still produce neighbourhoods good enough to preserve, rather than erode, the QKNN's detection rate.

Even at its best, however, the QKNN remains well below the stronger classical models of Table~\ref{tab:classical}: on the same ANOVA features, XGBoost and NN1 detect $31$ and $32$ suspicious accounts against the QKNN's $20$--$22$. The current fidelity-based representation leaves roughly half the suspicious accounts unflagged, which motivates learning the feature map rather than fixing it, a direction we return to in Sec.~\ref{sec:conclusion}.

\subsection{QSVM Results}
\label{sec:qsvm-results}
All QSVM results share a single kernel configuration: the one-body projected kernel of Eq.~\eqref{eq:qsvm_onebody} on $n = 20$ qubits encoding the $d = 20$ features of the ANOVA and PCA representations via the second-order Pauli-$ZZ$ feature map of Sec.~\ref{sec:methods:qsvm:featuremap}, with $d_m = 2$ repetitions, pairwise (linear) entanglement and $N_{\text{shots}} = 4096$, under the noise models of \texttt{FakeFez} and \texttt{FakeGuadalupeV2}. The precomputed kernel is fitted with the standard dual solver of Sec.~\ref{sec:methods:qsvm:comparison}. Table~\ref{tab:qsvm_results} reports each representation and noise model at the default threshold and at each thresholding criterion.
\begin{table}[H]
\centering
\caption{One-body projected-kernel QSVM on the $200$ test accounts ($40$ suspicious, $160$ benign), for each representation under the noise models of \texttt{FakeFez} and \texttt{FakeGuadalupeV2}. ``$\tau=0$'' is the standard SVM decision rule; the other rows are the threshold-changing criteria of Sec.~\ref{sec:methods:qsvm:comparison}, with thresholds selected on the test split. The precision floors $p_{\min}=0.55$ and $0.60$ gave identical results and are shown as one row. FPR is the percentage of the $160$ benign accounts that are flagged; TP, FN and FP are counts. Rows marked $\dagger$ flag every test account. The best F1 within each block is in bold. Precision, recall and F1 are for the suspicious class; all rates are in \%.}
\label{tab:qsvm_results}
\resizebox{\textwidth}{!}{%
\begin{tabular}{lccccccc}
\toprule
\textbf{Operating point} & \textbf{Precision (\%)} & \textbf{Recall (\%)} & \textbf{F1 (\%)} & \textbf{FPR (\%)} & \textbf{TP} & \textbf{FN} & \textbf{FP} \\
\midrule
\multicolumn{8}{l}{\textit{PCA, \texttt{FakeFez}}} \\
\quad $\tau = 0$ (standard) & 24 & 72 & 36 & 56.3 & 29 & 11 & 90 \\
\quad $F_\beta$, $\beta=1.25$ & 25 & 85 & \textbf{39} & 63.1 & 34 & 6 & 101 \\
\quad Recall max., $\mathrm{Precision}\ge0.55$ / $0.60$$^\dagger$ & 20 & 100 & 33 & 100.0 & 40 & 0 & 160 \\
\quad Precision max., $\mathrm{Recall}\ge0.70$ & 25 & 85 & \textbf{39} & 63.1 & 34 & 6 & 101 \\
\quad Precision max., $\mathrm{Recall}\ge0.75$ & 25 & 85 & \textbf{39} & 63.1 & 34 & 6 & 101 \\
\midrule
\multicolumn{8}{l}{\textit{PCA, \texttt{FakeGuadalupeV2}}} \\
\quad $\tau = 0$ (standard) & 23 & 62 & 34 & 51.3 & 25 & 15 & 82 \\
\quad $F_\beta$, $\beta=1.25$ & 24 & 85 & 37 & 67.5 & 34 & 6 & 108 \\
\quad Recall max., $\mathrm{Precision}\ge0.55$ / $0.60$$^\dagger$ & 20 & 100 & 33 & 100.0 & 40 & 0 & 160 \\
\quad Precision max., $\mathrm{Recall}\ge0.70$ & 25 & 78 & \textbf{38} & 58.1 & 31 & 9 & 93 \\
\quad Precision max., $\mathrm{Recall}\ge0.75$ & 25 & 78 & \textbf{38} & 58.1 & 31 & 9 & 93 \\
\midrule
\multicolumn{8}{l}{\textit{ANOVA, \texttt{FakeFez}}} \\
\quad $\tau = 0$ (standard) & 20 & 7 & 11 & 7.5 & 3 & 37 & 12 \\
\quad $F_\beta$, $\beta=1.25$ & 30 & 62 & \textbf{41} & 35.6 & 25 & 15 & 57 \\
\quad Recall max., $\mathrm{Precision}\ge0.55$ / $0.60$ & 60 & 7 & 13 & 1.3 & 3 & 37 & 2 \\
\quad Precision max., $\mathrm{Recall}\ge0.70$ & 24 & 70 & 36 & 54.4 & 28 & 12 & 87 \\
\quad Precision max., $\mathrm{Recall}\ge0.75$ & 23 & 78 & 36 & 64.4 & 31 & 9 & 103 \\
\midrule
\multicolumn{8}{l}{\textit{ANOVA, \texttt{FakeGuadalupeV2}}} \\
\quad $\tau = 0$ (standard) & 20 & 7 & 11 & 7.5 & 3 & 37 & 12 \\
\quad $F_\beta$, $\beta=1.25$ & 24 & 93 & \textbf{38} & 75.0 & 37 & 3 & 120 \\
\quad Recall max., $\mathrm{Precision}\ge0.55$ / $0.60$ & 67 & 5 & 9 & 0.6 & 2 & 38 & 1 \\
\quad Precision max., $\mathrm{Recall}\ge0.70$ & 24 & 93 & \textbf{38} & 75.0 & 37 & 3 & 120 \\
\quad Precision max., $\mathrm{Recall}\ge0.75$ & 24 & 93 & \textbf{38} & 75.0 & 37 & 3 & 120 \\
\bottomrule
\end{tabular}}
\end{table}
No operating point of the QSVM exceeds $41\%$ F1 on either representation or backend. At the standard threshold PCA is the stronger representation ($34$--$36\%$), whereas ANOVA nearly collapses: only $3$ of $40$ suspicious accounts are flagged, an F1 of $11\%$. ANOVA, the best representation for the QKNN, the HQNN and the classical tree and network models, is therefore the worst for the QSVM at its default threshold, and it is recovered only by moving the threshold: under $F_\beta$ maximisation on \texttt{FakeFez}, ANOVA reaches $41\%$, the best QSVM result.

The thresholding criteria expose the shape of the kernel's score distribution more than they improve on it. Every criterion that trades precision for recall lands in the same narrow band, precision of $23$--$30\%$ against a $20\%$ prevalence, at an FPR of $35.6$--$75.0\%$, that is, flagging between a third and three quarters of all benign accounts. The precision floors are the clearest case: at $p_{\min}=0.55$ or $0.60$, no threshold on PCA meets the floor and the procedure returns the degenerate flag-everything operating point, while on ANOVA it meets the floor only by flagging $3$ to $5$ accounts (FPR of $0.6$--$1.3\%$). The kernel thus does not rank any sizeable group of suspicious accounts clearly above the benign ones; the suspicious accounts it does identify are mixed with many more benign accounts at similar scores.

The two noise models agree exactly at the standard threshold on ANOVA and differ by $4$ detected accounts on PCA ($29$ against $25$), but differ more after thresholding, most visibly on ANOVA under $F_\beta$ ($62\%$ recall on \texttt{FakeFez} against $93\%$ on \texttt{FakeGuadalupeV2}). Because the threshold is re-selected for each backend, this sensitivity mainly reflects how flat the score distribution is around the chosen operating point rather than a large difference in the quality of the two kernels: F1 at those two points differs by only $3$ points ($41\%$ against $38\%$).

\subsection{QNN Results}

\label{sec:qnn-results}
All variational parameters were optimized with COBYLA (Constrained Optimization By Linear Approximation), a derivative-free trust-region method that builds a linear interpolation model of the objective from a simplex of sampled points. The choice is dictated by the cost of the circuit evaluation rather than the asymptotic convergence rate. A quasi-Newton method such as L-BFGS-B requires the gradient at every iterate, obtained by the parameter-shift rule, which costs $2P$ circuit evaluations per gradient for $P$ variational parameters, whereas COBYLA advances on a single objective evaluation per iterate once its initial simplex is built. For the parameter counts considered here, COBYLA therefore reaches a comparable objective value in substantially fewer circuit executions, which dominates wall-clock time. The trade-off is that COBYLA is a local method with no convergence guarantee under stochastic objectives, and its performance degrades as $P$ grows.

The QNN was trained for $800$ COBYLA iterations on the ideal simulator, on the ANOVA and PCA representations, under each of the four losses of Sec.~\ref{sec:loss}. The training loss had not fully converged at that point, but training the $20$-qubit circuit on $800$ accounts is computationally demanding, and longer runs did not improve test performance, so training was stopped at $800$ iterations. Table~\ref{tab:qnn_results} reports the best loss at each operating point.
\begin{table}[H]
\centering
\caption{QNN ($20$ qubits, ideal simulator) on the $200$ test accounts ($40$ suspicious, $160$ benign). Operating points as in Sec.~\ref{sec:postprocessing}. Precision, recall, F1 and FPR are in \%, for the suspicious class; TP, FN and FP are counts.}
\label{tab:qnn_results}
\begingroup
\small
\setlength{\tabcolsep}{4pt}
\begin{tabular}{llcccccccc}
\toprule
\textbf{Features} & \textbf{Op.\ point} & \textbf{Loss} & \textbf{Prec.} & \textbf{Rec.} & \textbf{F1} & \textbf{FPR} & \textbf{TP} & \textbf{FN} & \textbf{FP} \\
\midrule
\multirow{2}{*}{ANOVA} & FPR min.    & WMSE   & 28.6 & 55.0 & 37.6 & 34.4 & 22 & 18 & 55  \\
                       & Recall max. & WBCE   & 27.9 & 85.0 & \textbf{42.0} & 55.0 & 34 & 6  & 88  \\
\addlinespace
\multirow{2}{*}{PCA}   & FPR min.    & WHINGE & 26.0 & 50.0 & 34.2 & 35.6 & 20 & 20 & 57  \\
                       & Recall max. & WHINGE & 21.7 & 70.0 & 33.1 & 63.1 & 28 & 12 & 101 \\
\bottomrule
\end{tabular}
\endgroup
\end{table}
The QNN's best result is $42.0\%$ F1, on ANOVA at the recall-maximization point: it detects $34$ of the $40$ suspicious accounts, but only by flagging $88$ of the $160$ benign ones (FPR $55\%$), so fewer than three in ten of its alerts are genuine. At the tighter operating point it still flags a third of the benign accounts ($55$, FPR $34.4\%$) while detecting $22$. As for the other quantum models, ANOVA is the better representation, by $3$--$9$ points of F1. The QNN therefore sits in the same band as the QKNN and the QSVM, roughly $20$ points of F1 below the best classical models, and its precision of $22$--$29\%$ is only modestly above the $20\%$ of a rule that flags every account. A likely contributor is optimization rather than expressivity: the $80$ parameters of the $20$-qubit RealAmplitudes ansatz are trained by a derivative-free method that had not converged after $800$ iterations, and the fixed feature map gives the optimizer no way to reshape how accounts are embedded.

\subsection{HQNN Results}
\label{sec:hqnn-results}
The HQNN of Sec.~\ref{sec:hqnn} was trained on the $800$ training accounts of each representation on an ideal simulator, for $300$ epochs, and scored on the same $200$ test accounts. Both of its thresholds were fixed on the training scores alone (Sec.~\ref{sec:postprocessing}) and applied unchanged to the test set. Table~\ref{tab:hqnn_results} reports both operating points; the best loss was WMSE on ANOVA and WBCE on PCA.
\begin{table}[H]
\centering
\caption{HQNN ($5$ qubits, ideal simulator) on the $200$ test accounts. FPR-minimization: threshold chosen on training scores for a false-positive-rate budget of $15\%$, then maximum recall. Recall-maximization: threshold chosen on training scores for a looser budget of about $25\%$, then maximum recall. Precision, recall, F1 and FPR are in \%, for the suspicious class; TP, FN and FP are counts.}
\label{tab:hqnn_results}
\begingroup
\small
\setlength{\tabcolsep}{4pt}
\begin{tabular}{llcccccccc}
\toprule
\textbf{Features} & \textbf{Op.\ point} & \textbf{Loss} & \textbf{Prec.} & \textbf{Rec.} & \textbf{F1} & \textbf{FPR} & \textbf{TP} & \textbf{FN} & \textbf{FP} \\
\midrule
\multirow{2}{*}{ANOVA} & FPR min.    & WMSE & \textbf{50.0} & 60.0 & 54.5 & 15.0 & 24 & 16 & 24 \\
                       & Recall max. & WMSE & 44.6 & 82.5 & \textbf{57.9} & 25.6 & 33 & 7  & 41 \\
\addlinespace
\multirow{2}{*}{PCA}   & FPR min.    & WBCE & 48.9 & 57.5 & 52.9 & 15.0 & 23 & 17 & 24 \\
                       & Recall max. & WBCE & 38.8 & 65.0 & 48.6 & 25.6 & 26 & 14 & 41 \\
\bottomrule
\end{tabular}
\endgroup
\end{table}
At the recall-maximization point on ANOVA, the HQNN detects $33$ of the $40$ suspicious accounts while flagging $41$ of the $160$ benign ones (F1 $57.9\%$). At the FPR-minimization point it flags only $24$ benign accounts ($15\%$) and detects $24$ suspicious ones, so exactly one alert in two is a genuine case (precision $50.0\%$). Its F1 of $48.6$--$57.9\%$ across the four operating points is the highest of any quantum or hybrid model in this study, and the threshold-free area under the precision--recall curve, $\approx0.51$ on ANOVA, is comparable to the average precision of the best classical model, ANOVA XGBoost ($0.505$, Table~\ref{tab:classical}), although the two estimators differ slightly.

\paragraph{Against the classical baselines.} On ANOVA features the HQNN's best F1 ($57.9\%$) is ahead of logistic regression ($41.3\%$) and NN3 ($52.4\%$) and behind NN1 ($59.8\%$) and XGBoost ($61.4\%$). Its recall-maximization point is very close to NN1's operating point: one more suspicious account detected ($33$ against $32$) at the cost of $6$ more false alarms ($41$ against $35$). Compared with XGBoost, the HQNN at its FPR-minimization threshold flags fewer benign accounts ($24$ against $30$) but detects $7$ fewer suspicious ones ($24$ against $31$). On PCA, the HQNN's FPR-minimization point ($52.9\%$) is above every classical PCA configuration, the best of which is logistic regression at $51.6\%$. These gaps of $1$--$4$ points of F1 are comparable to the seed-to-seed spread of the classical networks (Table~\ref{tab:classical}), so on this split the HQNN is competitive with, but not better than, the strongest classical models.

\paragraph{Against the QNN.} The HQNN achieves this with a five-qubit register, a quarter of the $20$ qubits required by the QKNN, the QSVM and the QNN, because the classical network absorbs the input dimension (Sec.~\ref{sec:hqnn}). At a similar detection rate on ANOVA ($33$ against $34$ of $40$ suspicious accounts), the HQNN raises fewer than half as many false alarms as the QNN ($41$ against $88$), and its F1 is $15.9$ points higher ($57.9$ against $42.0\%$). At the tighter operating point the HQNN both detects more ($24$ against $22$) and flags fewer than half as many benign accounts ($24$ against $55$). On our runs, the HQNN also trained in about a quarter of the QNN's wall-clock time, on the same machine. Among the quantum approaches evaluated here, the hybrid design is therefore clearly the most effective for this task.

\paragraph{Execution on quantum hardware.} The trained HQNN was also executed on the \texttt{ibm\_fez} superconducting processor, as a pure inference task with the network weights obtained on the ideal simulator held fixed. Table~\ref{tab:hqnn_hardware} reports the resulting test F1. On ANOVA the hardware F1 is $55.4\%$ at the FPR-minimization point and $59.8\%$ at the recall-maximization point, at or slightly above the simulated values and level with the classical NN1 ($59.8\%$). On PCA the recall-maximization point is again close to its simulated value ($51.2$ against $48.6\%$), while the FPR-minimization point drops by about $10$ points ($43.1$ against $52.9\%$). With $40$ suspicious test accounts, the ANOVA differences correspond to one or two accounts and should not be read as hardware improving the model; what the run does establish is that the five-qubit circuit survives transpilation and device noise with its detection performance broadly intact.
\begin{table}[H]
\centering
\caption{HQNN test F1 (\%) on the ideal simulator and on \texttt{ibm\_fez}, with network weights trained on the simulator and held fixed.}
\label{tab:hqnn_hardware}
\begin{tabular}{llcc}
\toprule
\textbf{Features} & \textbf{Op.\ point} & \textbf{Ideal simulator} & \textbf{\texttt{ibm\_fez}} \\
\midrule
ANOVA & FPR min.    & 54.5 & 55.4 \\
ANOVA & Recall max. & 57.9 & \textbf{59.8} \\
PCA   & FPR min.    & 52.9 & 43.1 \\
PCA   & Recall max. & 48.6 & 51.2 \\
\bottomrule
\end{tabular}
\end{table}

\subsection{Comparative Summary of Classical and Quantum Methods}
\label{sec:comparative-summary}
\begin{figure}[H]
\centering
\includegraphics[width=0.85\textwidth]{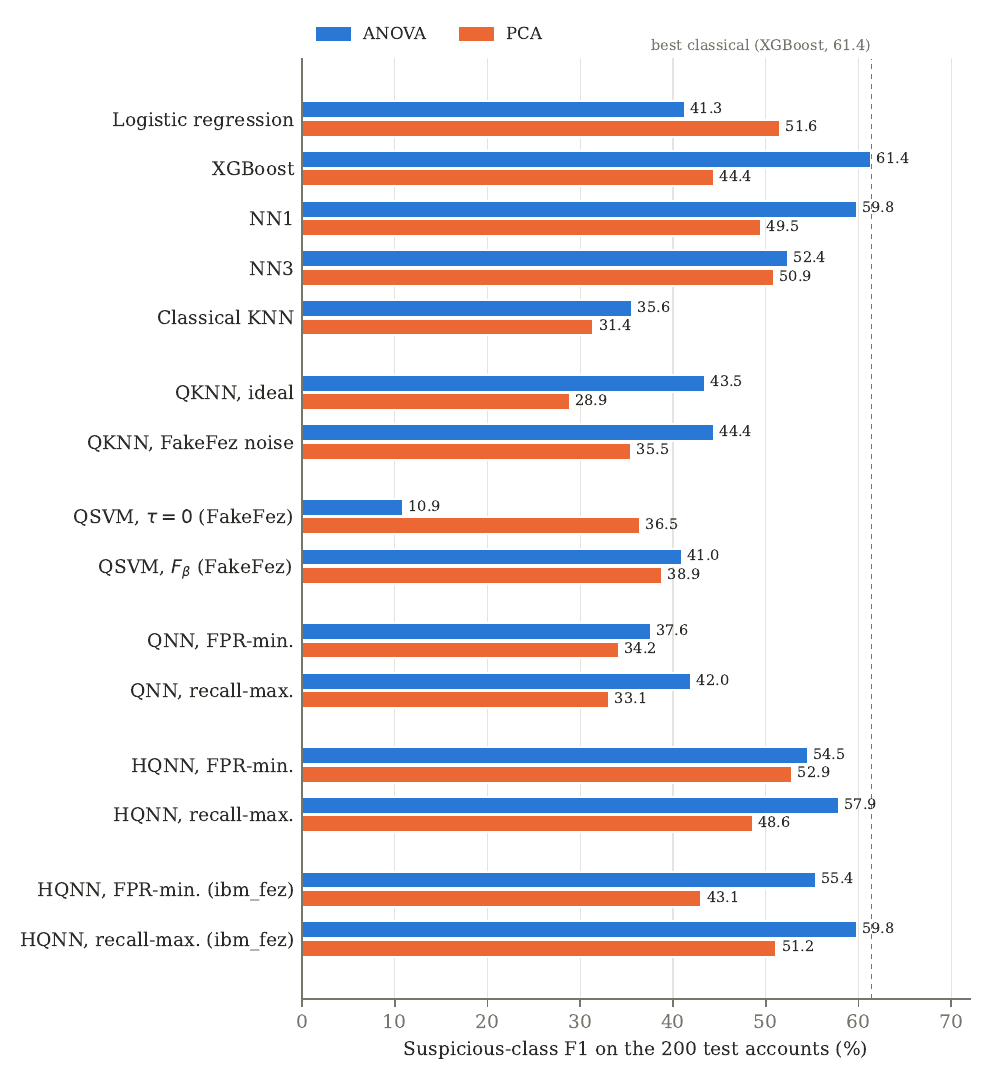}
\caption{Suspicious-class test F1 by method and representation (values as in Table~\ref{tab:method_comparison}). Bar colour marks the representation; the dashed line marks the best classical result, XGBoost on ANOVA ($61.4\%$).}
\label{fig:f1_comparison}
\end{figure}

\begin{figure}[H]
\centering
\includegraphics[width=0.85\textwidth]{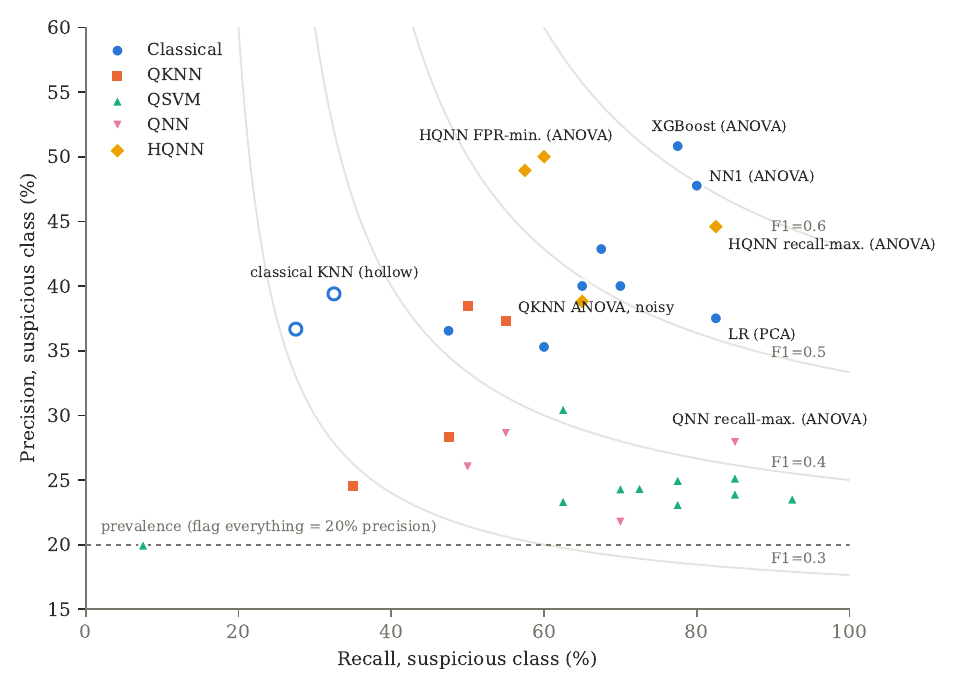}
\caption{Test precision against test recall for every simulated or classical operating point in Tables~\ref{tab:classical}--\ref{tab:hqnn_results}, except the QSVM precision-floor rows. Grey curves are lines of constant F1; the dashed line is the $20\%$ precision of a rule that flags every account. Marker shape and colour both identify the method family; hollow markers are the classical KNN control.}
\label{fig:pr_points}
\end{figure}

Table~\ref{tab:method_comparison} and Fig.~\ref{fig:f1_comparison} collect the test F1 of every method on every representation it was run on; Fig.~\ref{fig:pr_points} places the same operating points in precision--recall space.
\begin{table}[H]
\centering
\caption{Suspicious-class test F1 (\%) by method and representation, collected from Tables~\ref{tab:classical}--\ref{tab:hqnn_hardware}. ``Threshold'' gives how the operating point was chosen: OOF, maximum out-of-fold F1 on the training accounts; none, labels issued directly; $\tau{=}0$, default SVM rule; $F_\beta$, post hoc $F_\beta$ maximisation on the test split; FPR-min.\ and recall-max., false-positive-rate budgets of about $15\%$ and $25\%$ set on training scores. Best value in each column in bold.}
\label{tab:method_comparison}
\resizebox{\textwidth}{!}{%
\begin{tabular}{llcccc}
\toprule
\textbf{Method} & \textbf{Threshold} & \textbf{Qubits} & \textbf{Execution} & \textbf{ANOVA} & \textbf{PCA} \\
\midrule
Logistic regression & OOF & -- & classical & 41.3 & 51.6 \\
XGBoost & OOF & -- & classical & \textbf{61.4} & 44.4 \\
NN1 & OOF & -- & classical & 59.8 & 49.5 \\
NN3 & OOF & -- & classical & 52.4 & 50.9 \\
Classical KNN & none & -- & classical & 35.6 & 31.4 \\
\midrule
QKNN & none & 20 & ideal simulator & 43.5 & 28.9 \\
QKNN & none & 20 & \texttt{FakeFez} noise & 44.4 & 35.5 \\
QSVM (one-body PQK) & $\tau{=}0$ & 20 & \texttt{FakeFez} noise & 10.9 & 36.5 \\
QSVM (one-body PQK) & $F_\beta$ & 20 & \texttt{FakeFez} noise & 41.0 & 38.9 \\
QNN & FPR-min. & 20 & ideal simulator & 37.6 & 34.2 \\
QNN & recall-max. & 20 & ideal simulator & 42.0 & 33.1 \\
HQNN & FPR-min. & 5 & ideal simulator & 54.5 & \textbf{52.9} \\
HQNN & recall-max. & 5 & ideal simulator & 57.9 & 48.6 \\
HQNN & FPR-min. & 5 & \texttt{ibm\_fez} hardware & 55.4 & 43.1 \\
HQNN & recall-max. & 5 & \texttt{ibm\_fez} hardware & 59.8 & 51.2 \\
\bottomrule
\end{tabular}}
\end{table}
Three groups separate clearly in Fig.~\ref{fig:pr_points}. The classical models and the HQNN occupy the upper right, between the F1 $= 0.5$ and $0.6$ curves, with precision of $35$--$51\%$ at recall of $47$--$83\%$; the HQNN's FPR-minimization points reach the highest precision of any quantum or hybrid model ($49$--$50\%$). The QKNN sits lower and to the left, with its ANOVA configurations at precision near $38\%$ but recall of only $50$--$55\%$. The QNN and QSVM operating points lie in a band just above the prevalence line, at precision of $22$--$31\%$, wherever the threshold puts their recall; the one exception, the QSVM on ANOVA at its default threshold, sits on the prevalence line at $7.5\%$ recall. Ordered by best test F1, XGBoost ($61.4\%$) leads, followed by NN1 and the HQNN on hardware (both $59.8\%$), the HQNN on the ideal simulator ($57.9\%$), the remaining classical configurations ($41$--$52\%$), and then the QKNN ($44.4\%$), the QNN ($42.0\%$) and the QSVM ($41.0\%$). We return to this pattern in Sec.~\ref{sec:discussion}.
% =====================================================================
\newpage
\section{Discussion}
\label{sec:discussion}
\subsection{Comparable at Best, Not Superior}
On this synthetic pump-and-dump benchmark, with $20$ features per account and $40$ suspicious accounts in the test set, no quantum or hybrid model exceeds the best classical model. XGBoost on ANOVA features reaches $61.4\%$ F1 and the best quantum-hybrid model, the HQNN, $57.9\%$ on the ideal simulator and $59.8\%$ on \texttt{ibm\_fez}; these gaps are within the seed-to-seed spread of the classical networks, so the two are competitive, but no ordering in favour of the quantum model is supported. The other three quantum families are clearly behind: the QKNN by about $17$ points of F1, the QNN by about $19$ and the QSVM by about $20$.

What separates the HQNN from the other quantum models is where and how training happens. The QKNN's only adaptive elements are $k$ and its cluster routing, and the QSVM's kernel is fixed, with only the classical dual solver trained on it; within the QSVM, no choice of threshold recovers a separation that the fixed kernel does not already contain. The QNN does train its circuit, but its $80$ ansatz parameters sit behind a fixed $20$-qubit feature map and are optimized by a derivative-free method that had not converged after $800$ iterations. The HQNN instead trains $948$ classical parameters by gradient descent, through the circuit, and in doing so learns how each account is embedded. The evidence suggests that, on this task, a quantum circuit contributes usefully only when it is paired with an efficient optimization loop that adapts the embedding itself to the labels.

This also limits what the HQNN result shows. Its parameter count is of the same order as that of NN1 ($705$ parameters at the selected width of $32$), and NN1 on the same features reaches a similar F1. The comparison therefore does not show that the five-qubit circuit contributes anything that a classical layer of similar size would not; establishing that requires an ablation in which the circuit is replaced by a classical layer with the same number of outputs, trained identically, which we leave to future work. What the result does show is that a hybrid model on five qubits reaches the same performance band as the strongest classical models, with thresholds fixed on training data alone, on real quantum hardware as well as in simulation, while using a quarter of the qubits of every other quantum model in this study.

\subsection{Operational Significance}
\label{sec:operational}
The comparison above is between models; the comparison that matters to a compliance team is with current practice. AML and trade-surveillance monitoring today is overwhelmingly manual: rule-based systems raise alerts, and every alert is reviewed by a human analyst. Industry estimates put the share of those alerts that turn out to be false positives above $95\%$~\cite{han2021,lexisnexis2022}, so fewer than one alert in twenty leads to a genuine finding, and most analyst time is spent clearing legitimate activity.

Against that baseline, the operating points reported here represent a substantial change. At its FPR-minimization point, the HQNN flags $15\%$ of benign accounts and half of its alerts are genuinely suspicious (precision $50.0\%$), while still detecting $60\%$ of the suspicious accounts; the best classical model, XGBoost, reaches a similar precision ($50.8\%$) at higher recall. Even an F1 of around $50\%$, which every HQNN configuration on ANOVA attains, corresponds to a level of alert quality roughly an order of magnitude better than the one-in-twenty yield of rule-based monitoring. Two cautions apply. First, precision depends on prevalence: our benchmark is $20\%$ suspicious, far above the rate in production data, and at a lower prevalence the same false-positive rate and recall would yield a lower precision; the false-positive rate and recall themselves carry over more directly. Second, these are results on synthetic data. The comparison with current practice is therefore indicative rather than a measured improvement, but it shows that models in this performance band, classical or hybrid, would materially reduce the manual review burden that dominates AML operations today.

\subsection{Representation Matters as Much as the Model}
The choice between ANOVA and PCA changes F1 by as much as any choice of model. ANOVA, which keeps $20$ of the original features rather than combining them, is the best representation for XGBoost, both classical networks, the QKNN (by $15$ points on the ideal simulator), the QNN, the HQNN at its best operating point and, at a tuned threshold, the QSVM. PCA is the better representation only for logistic regression and for the QSVM at its default threshold. The QSVM's near-collapse on ANOVA at its default threshold ($3$ of $40$ suspicious accounts detected), where every other model does best, indicates that the one-body projection discards much of the information that the selected features carry. A plausible mechanism is that projecting the $20$-qubit state onto one-body reduced density matrices retains little of the pairwise structure that the $ZZ$ feature map encodes; testing this would require comparing against the full fidelity kernel and against projections onto two-body marginals on the same data.

\subsection{Noise and Hardware}
Every method exposed to noise tolerated it. The QKNN's recall rose under the \texttt{FakeFez} noise model on both representations; the QSVM's results under two different noise models agree exactly at the standard threshold on ANOVA and differ by four detected accounts on PCA; and the HQNN, executed on \texttt{ibm\_fez} with weights trained in simulation, kept or slightly exceeded its simulated F1 at three of its four operating points. The fake-backend results do not by themselves establish robustness on physical hardware, and the QKNN comparison is confounded by a change of shot budget; the HQNN run, by contrast, is on a real device, but it is a single inference-only run. Together they indicate that, at the circuit depths used here, noise is not the factor limiting any of the methods; the representation and the training procedure are. The HQNN's five-qubit, shallow circuit is also the easiest of the four quantum models to run on current hardware, a practical advantage that grows with the gap between required and available qubit counts.

\subsection{Class Imbalance and False Negatives}
Because a missed suspicious account is more costly in a deployed AML system than an extra alert, recall at an acceptable false-alarm burden is the quantity of most practical interest. The HQNN's operating points illustrate how this can be controlled directly through a false-positive-rate budget set on training data, which unlike a precision target does not depend on prevalence. By contrast, the QSVM's precision floors, also intended to control the alert burden, were unattainable on the PCA representation, and the QKNN offers no threshold through which the trade-off can be tuned. Extending threshold selection to the QKNN's vote margin, and fixing the QSVM's thresholds on training data, are the natural steps towards putting all methods on the same operating-point footing.

\subsection{Limitations and Threats to Validity}
Six limitations qualify the comparison. First, the methods do not share one operating-point convention: the classical models use out-of-fold-F1 thresholds, the QNN and HQNN false-positive-rate budgets on training scores, the QKNN none, and the QSVM post hoc criteria selected on the test split, so Table~\ref{tab:method_comparison} compares each method at its own natural operating point rather than at matched precision or recall; because its thresholds are chosen on the same accounts on which they are scored, the QSVM's threshold-changing results are an optimistic upper bound, whereas its standard ($\tau=0$) results involve no such selection. Second, the test set contains only $40$ suspicious accounts, so one account moves recall by $2.5$ points, and no confidence intervals or paired significance tests are reported because the independence of accounts belonging to the same pump-and-dump event has not been verified; differences of a few points of F1 should not be read as rankings. Third, the $20$-feature representations were fitted once on all $800$ training accounts, so the cross-validation folds used to select the classical models' hyperparameters and thresholds were not fully isolated from the feature selection or projection, and the classical out-of-fold scores are somewhat optimistic; the test-set results are not affected. Fourth, the quantum results come from a single training run each, whereas the classical results include a ten-seed sensitivity analysis. Fifth, only the HQNN was executed on physical quantum hardware, as a single inference-only run with weights trained in simulation; the QKNN and QSVM were run under simulated noise and the QNN on an ideal simulator, and hardware trainability was not tested for any model. Sixth, the benchmark is synthetic and its $20\%$ suspicious rate is far above the well-below-$1\%$ rate of production AML systems; neither the absolute levels nor the ranking should be assumed to transfer unchanged to real trade-surveillance data.
% =====================================================================
\newpage
\section{Conclusion}
\label{sec:conclusion}
This work benchmarked four quantum or hybrid quantum--classical classifier families, a clustering-based fidelity QKNN, a one-body projected-kernel QSVM, a fixed-feature-map QNN and a hybrid quantum neural network, against logistic regression, XGBoost, two neural networks and a classical KNN on a synthetic pump-and-dump detection task, with every model trained on the same $800$ accounts, tested on the same $200$, and fed the same two $20$-feature representations. The strongest model was classical, XGBoost on ANOVA features at $61.4\%$ F1. The HQNN, a five-qubit circuit whose every angle is generated by a small classical network, was the strongest quantum-hybrid model: $54.5$--$57.9\%$ F1 on ANOVA in simulation and $55.4$--$59.8\%$ on the \texttt{ibm\_fez} processor, ahead of six of the eight classical configurations, and well ahead of the $20$-qubit QNN ($42.0\%$), which it beat with fewer than half the false alarms, a quarter of the qubits and about a quarter of the training time. The QKNN improved on a classical $k$-nearest-neighbours control and kept its performance under simulated device noise, and the projected-kernel QSVM did not exceed $41\%$ F1 at any operating point. We do not report a quantum advantage.

The practical point is nonetheless significant. AML and trade-surveillance review is still largely manual, with more than nineteen in twenty rule-based alerts turning out to be false. A model that makes one alert in two a genuine case, as the HQNN and the best classical models do here, would reduce that burden by roughly an order of magnitude, subject to the prevalence caveats of Sec.~\ref{sec:operational}. That a five-qubit hybrid model reaches this level, on real quantum hardware, is the main result of this comparison.

The most direct next steps follow from these results: an ablation that replaces the HQNN's circuit with a classical layer of the same size, to establish whether the quantum layer contributes; richer HQNN circuits, in particular the multi-qubit data re-uploading circuit, with the network setting both single- and two-qubit gates; a learned rather than fixed feature map for the QKNN, for example by freezing a quantum embedding trained as a neural quantum kernel~\cite{rodriguezgrasa2025neural} and using it to build the fidelity kernel and clusters; comparing the one-body QSVM kernel with the full fidelity kernel and with two-body projections on the same data; matched operating-point criteria across all methods; repeated quantum runs over several seeds; training, not only inference, on hardware; and a repeat of the comparison at a production-realistic suspicious rate.
% =====================================================================
\newpage
\appendix
\section{An Ising-Hamiltonian-Motivated Loss Function}
\label{sec:ising-loss}
To ground the training objective in a physically motivated energy functional, we derive a classification loss from the ground-state energetics of a diagonal Ising Hamiltonian evaluated on a fixed reference state.
\subsection{The Hamiltonian}
Consider $N$ qubits arranged on an open one-dimensional chain, so that there are $N-1$ nearest-neighbor bonds $\langle i,j\rangle$. We take the ferromagnetic, longitudinal-field Ising Hamiltonian
\begin{equation}
H = -\sum_{\langle i,j\rangle} Z_i Z_j \;-\; h \sum_{i=1}^{N} Z_i ,
\label{eq:ising-general}
\end{equation}
where $Z_i$ is the Pauli-$Z$ operator on qubit $i$ ($Z_i\ket{0}_i = +\ket{0}_i$, $Z_i\ket{1}_i = -\ket{1}_i$) and $h$ is a field strength. Every term in \eqref{eq:ising-general} is diagonal in the computational basis, so $\langle H\rangle$ can be read off directly from measured bit statistics without further processing.
To couple $H$ to a binary classification task, we set the field strength to the product of the true label $Y\in\{-1,+1\}$ and the model's continuous prediction $\hat Y \in [-1,1]$ (e.g.\ a Pauli-$Z$ expectation value read out from a variational quantum circuit):
\begin{equation}
H = -\sum_{\langle i,j\rangle} Z_i Z_j \;-\; Y\hat Y \sum_{i=1}^{N} Z_i .
\label{eq:ising-parameterized}
\end{equation}
We evaluate $\langle H\rangle$ on the fixed reference state $\ket{0}^{\otimes N}$, so that all dependence on the classification outcome enters solely through $Y\hat{Y}$.
\subsection{Ground-State Case Analysis}
In $\ket{0}^{\otimes N}$ every qubit sits in the $Z_i=+1$ eigenstate, so
\begin{equation}
\Big\langle \sum_{\langle i,j\rangle} Z_i Z_j \Big\rangle_{\ket{0}^{\otimes N}} = N-1, \qquad \Big\langle \sum_{i=1}^{N} Z_i \Big\rangle_{\ket{0}^{\otimes N}} = N,
\end{equation}
giving
\begin{equation}
\langle H \rangle_{\ket{0}^{\otimes N}} = -(N-1) - Y\hat{Y}\,N .
\label{eq:H-expectation}
\end{equation}
Two limiting cases characterize $Y\hat Y$ as an agreement indicator between the true label and the prediction:
\begin{itemize}
  \item \textbf{Correct, fully confident} ($Y\hat Y = +1$):
  \begin{equation}
  \langle H\rangle_{\ket{0}^{\otimes N}} = -(N-1) - N = -(2N-1),
  \end{equation}
  the \emph{minimum} energy attainable by \eqref{eq:H-expectation}.
  \item \textbf{Incorrect, fully confident} ($Y\hat Y = -1$):
  \begin{equation}
  \langle H\rangle_{\ket{0}^{\otimes N}} = -(N-1) + N = 1,
  \end{equation}
  higher than the correct case by $2N$.
\end{itemize}
This is exactly the behavior required of a training objective: agreement between $Y$ and $\hat Y$ is energetically favored, and disagreement is penalized.
\subsection{Loss Function}
Equation~\eqref{eq:H-expectation} is not itself a usable loss: it is non-zero even for a perfect prediction. We fix this by subtracting off the ground-state energy $-(2N-1)$ found above, i.e.\ shifting $\langle H\rangle$ by $+(2N-1)$, so that the loss vanishes exactly when the prediction is correct and fully confident:
\begin{align}
\mathcal{L} &\equiv \langle H \rangle_{\ket{0}^{\otimes N}} + (2N-1) \notag \\
&= \underbrace{-(N-1) - Y\hat Y\,N}_{-N(1+Y\hat Y)+1} \;+\; (2N-1) \notag \\
&= N - N Y\hat Y .
\end{align}
For a single training example this gives
\begin{equation}
\boxed{\mathcal{L}(Y,\hat Y) = N\left(1 - Y\hat{Y}\right)} ,
\label{eq:loss-single}
\end{equation}
and, averaged over a batch of $M$ samples,
\begin{equation}
\boxed{\mathcal{L} = \frac{1}{M}\sum_{k=1}^{M} N\left(1 - Y_k\hat{Y}_k\right)},
\label{eq:loss-batch}
\end{equation}
which resembles eq. \eqref{eq:mhinge}. Since $Y_k \in \{-1,+1\}$ and $\hat Y_k \in [-1,1]$, this loss is an energy \emph{gap above the Ising ground state}: it satisfies $\mathcal{L} \in [0, 2N]$, vanishing only when every prediction is both correct and maximally confident, and growing linearly with the disagreement $Y_k\hat Y_k$ rather than saturating, unlike a hinge loss.

\section*{Acknowledgments}
This work was supported through the partnership between Vanguard, Capgemini, and The Washington Institute for STEM, Entrepreneurship, and Research (WISER), which provided the fellowship enabling this research.

\section*{Abbreviations}
\noindent The following abbreviations are used in this manuscript:\\[4pt]
\noindent
\begin{tabular}{@{}ll}
AML & Anti-money laundering\\
NISQ & Noisy intermediate-scale quantum\\
QKNN & Quantum $k$-nearest neighbours\\
QNN & Quantum neural network\\
QSVM & Quantum support vector machine\\
HQNN & Hybrid quantum neural network\\
PQK & Projected quantum kernel\\
ANOVA & Analysis of variance\\
PCA & Principal component analysis\\
UMAP & Uniform manifold approximation and projection\\
FPR & False-positive rate\\
OOF & Out-of-fold\\
\end{tabular}

\bibliographystyle{unsrtnat}
\bibliography{references,ref-qnn,ref-qsvm}

\end{document}